\pdfoutput=1
\RequirePackage{fix-cm}
\documentclass[smallextended]{svjour3}    
\smartqed  
\usepackage{graphicx}
\usepackage{float}
\usepackage{mathptmx}      
\usepackage{amsmath}
\usepackage{siunitx}
\usepackage[title]{appendix}
\usepackage{circuitikz}

\journalname{Experimental Astronomy}

\begin{document}

\title{SARAS 3 CD/EoR Radiometer: Design and Performance of the Receiver}

\titlerunning{SARAS 3 radiometer receiver} 

\author{Jishnu Nambissan T.\textsuperscript{1,2}  \and Ravi Subrahmanyan\textsuperscript{1} \and R. Somashekar\textsuperscript{1} \and N. Udaya Shankar\textsuperscript{1} \and Saurabh Singh\textsuperscript{1,3,4}  \and A. Raghunathan\textsuperscript{1} \and B.S. Girish\textsuperscript{1}  \and K.S. Srivani\textsuperscript{1} \and Mayuri Sathyanarayana Rao\textsuperscript{1,5}}

\institute{Jishnu Nambissan T. \at
              \email{jishnu@rri.res.in} 
              \and 
              \at \textsuperscript{1} Raman Research Institute, C V Raman Avenue, Sadashivanagar, Bangalore 560080, India
              \at \textsuperscript{2} International Centre for Radio Astronomy Research, Curtin University, Bentley, WA 6102, Australia
              \at \textsuperscript{3} McGill Space Institute, McGill University, 3550 rue University, Montr\'eal, QC H3A 2A7, Canada
              \at \textsuperscript{4} Department of Physics, McGill University, 3600 rue University, Montr\'eal, QC H3A 2T8, Canada
              \at \textsuperscript{5} Physics Division, Lawrence Berkeley National Lab, 1 Cyclotron Road, Berkeley, CA 94720, USA
              }

\date{Received: 01 June 2020 / Accepted: 01 July 2020}
\maketitle

\begin{abstract}
SARAS is an ongoing experiment aiming to detect the redshifted global 21-cm signal expected from Cosmic Dawn (CD) 
and the Epoch of Reionization (EoR). Standard cosmological models predict the signal to be present in the redshift range $z \sim $6--35, corresponding to a frequency range 40--200~MHz, as a spectral distortion of amplitude 20--200~mK in the 3~K cosmic microwave background. Since the signal might span multiple octaves in frequency, and this frequency range is dominated by strong terrestrial Radio Frequency Interference (RFI) and astrophysical foregrounds of Galactic and Extragalactic origin that are several orders of magnitude greater in brightness temperature, design of a radiometer for measurement of this faint signal is a challenging task. It is critical that the instrumental systematics do not result in additive or multiplicative confusing spectral structures in the measured sky spectrum and thus preclude detection of the weak 21-cm signal. Here we present the system design of the SARAS~3 version of the receiver.  New features in the evolved design include Dicke switching, double differencing and optical isolation for improved accuracy in calibration and rejection of additive and multiplicative systematics.  We derive and present the measurement equations for the SARAS~3 receiver configuration and calibration scheme, and provide results of laboratory tests performed using various precision terminations that qualify the performance of the radiometer receiver for the science goal.


\end{abstract}

\section{Introduction}
\label{intro}
Following cosmological recombination of the primordial hydrogen and helium, the Dark Ages is expected to have ended with the formation of the first 
stars in the first ultra-faint galaxies, which lit up the universe.  During this transformational epoch spanning redshifts $z \sim 6$--35, 
commonly referred to as the Cosmic Dawn (CD) and the Epoch of Reionization (EoR), 
the baryons in the universe transitioned from being mostly neutral during the Dark Ages to being almost completely ionized by the end of EoR \cite{doi:10.1111/j.1365-2966.2005.09485.x} \cite{1999A&A...345..380S}. 
However, the nature of the sources driving this transition, the timing of events, and the physical---light-matter and gravitational hydrodynamic---processes that govern the evolving gas transition are poorly understood. 
A key reason for the uncertainty is the lack of observational constraints.

In this context, it has been recognised that the redshifted 21-cm signal from neutral hydrogen at those epochs could act as a direct probe to trace the evolving gas properties of the Intergalactic Medium (IGM) in the Dark Ages, Cosmic Dawn and Reionization.  
Specifically, the sky-averaged or global component of this 21-cm signal has been shown to be an extremely powerful tool \cite{doi:10.1093/mnras/stx2065}\cite{PhysRevD.82.023006}; therefore, several experiments are currently underway to detect this signal.  SARAS (Shaped Antenna measurement of the background RAdio Spectrum) is a spectral radiometer experiment aiming to measure the spectrum of the global 21~cm signal.  
Apart from SARAS, which is the subject of this manuscript, other ongoing 
experiments include EDGES\cite{Bowman2018}, SCI-HI\cite{2014ApJ...782L...9V}, BIGHORNS\cite{2015PASA...32....4S}, PRIZM and LEDA\cite{doi:10.1093/mnras/sty1244}.

The first version, SARAS~1 \cite{2013ExA....36..319P}, operated in the band 87.5--175~MHz and employed a frequency-independent fat-dipole antenna above ferrite-tile absorbers as the sensor of the sky radiation.  The signal received at the antenna terminal was split as the first stage of analog signal processing, propagated through independent receiver chains, and finally digitised and correlated using a 1024-channel digital cross-correlation spectrometer to measure the sky spectrum.  
Adoption of a correlation radiometer concept, usage of optical fibre links and phase switching were aimed at suppression of internal systematics. 
SARAS~1 provided an improved calibration for the 150~MHz all-sky map of Landecker \& Wielebinski \cite{0004-637X-801-2-138}; 
however, RFI, non-smooth spectral behaviour of the balun, limited absorbtivity of the ferrite-tile ground and spectral confusion arising from multi-path interference 
within the signal path limited the sensitivity.

SARAS~2 \cite{Singh2018}, the second version of SARAS, employed a 8192 channel cross correlation spectrometer along with an electrically short spherical monopole antenna.  Evolution from dipole to monopole antenna allowed the elimination of balun and also substantial reduction in signal transmission length between antenna and receiver.  As a consequence, confusion from multi-path interference within the signal path was reduced and hence spectral smoothness in the instrument response was improved. Additionally, the use of a monopole antenna instead of a dipole avoided beam chromaticity that arises from reflections off the ground plane beneath the antenna,  thereby reducing spectral confusion due to mode coupling of spatial structures in the sky foreground into frequency structure. Based on 63~hr of data collected at the Timbaktu Collective in Southern India, SARAS~2 placed constraints on the global 21~cm signal, 
disfavouring models with low X-ray heating and rapid reionization \cite{2041-8205-845-2-L12} \cite{0004-637X-858-1-54}.  The poor efficiency of the short monopole, which worsened at low frequencies, limited the scientifically useful band of SARAS~2 to 110--200~MHz; additionally, the sensitivity suffered due to the excess system temperature of the adopted correlation receiver configuration.

In this paper, we describe the present improved version of SARAS, referred to as SARAS~3, which has an architecture  different from earlier versions.
First, introducing a double differencing strategy improves calibration and cancellation of unwanted additive features in the spectra.  
Second, in order to reduce the additive noise from the splitter of the correlation spectrometer, splitting of the signal received by the antenna is carried out post signal amplification.  The receiver system is designed to operate over frequencies 40--230~MHz, and intended to be used with scaled conical monopole antennas that operate in octave bands and cover the range in staggered bands.

In the following sections, we describe the various considerations that have gone into the design of SARAS~3,
emphasising improvements in design and performance. In Section~\ref{sec:motiv}, we discuss the motivation for the SARAS~3 version of the receiver and in Section~\ref{sec:sys_overview}, we present an overview of the system. 
In Section~\ref{sec:meas_eq}, we derive the measurement equation. The implementation of the analog receiver is described in Section~\ref{sec:implementation}. Section~\ref{sec:sensit} presents the measurement sensitivity 
wherein we also demonstrate that SARAS~3 is capable of detection of EoR signals with amplitudes and spectral complexity predicted in standard models. 
Laboratory tests that qualify receiver performance are discussed in Section~\ref{sec:lab} to demonstrate adequate control of internal systematics.

\section{Motivation for the SARAS~3 receiver}
\label{sec:motiv}
SARAS~2, the second version of SARAS, was a cross correlation differential spectrometer \cite{Singh2018}. The spectra measured were differential measurements made between the antenna and an internal reference load. The antenna was a spherically shaped monopole antenna of height 33 cm over a ground disc of radius 43.5 cm.  The antenna was designed to be electrically short at all frequencies below 200~MHz thus capable of observing over the entire CD/EoR band with a fairly frequency independent beam pattern; however, the downside was the substantial loss in radiative and reflection efficiencies at longer wavelengths.

The first element in the front-end electronics was a four-port cross over switch with the antenna connected to one input port and a calibration noise source connected to the second port through a series of attenuators, which also served as the internal reference load. The outputs of the cross-over switch were connected to a $180^{\circ}$ hybrid, which produced sum and differences of the input antenna and reference signals. These sum and difference signals were then amplified and transmitted via a pair of 100~m radio frequency over fibre (RFoF) links to the back-end electronics. The front-end electronics was in a shielded enclosure underground beneath the antenna.  At the back-end receiver, the pair of optical signals were demodulated to electrical signals and then processed by two identical arms of signal processing electronics consisting of amplifiers and band limiting filters. The passband was band-limited to 40--230~MHz where the CD/EoR signal is predicted to be. The processed signals were sampled, digitised and Fourier transformed using a 8192-point Fast Fourier Transform (FFT) algorithm on a Virtex-6 FPGA, then cross correlated to yield the correlation spectra. Bandpass as well as absolute calibration of the spectra was implemented by switching on and off the noise source, which introduced a step of calibrated noise temperature in the measured spectra. For details of the SARAS~2 system and its performance, we refer the reader to \cite{Singh2018}.

The SARAS design philosophy has been to purpose design all multiplicative gains and additive systematics, which survive calibration with the noise injection, to be spectrally smooth. We define smoothness as given by the class of polynomials having no zero crossings in second and higher-order derivatives; such polynomials are called maximally smooth (MS) polynomials \cite{2017ApJ...840...33S}.  SARAS~2 included a 180-degree hybrid in the front-end receiver that served to split the signal from the antenna between the two arms of the correlation spectrometer.  This hybrid impressed frequency-dependent loss as well as additives on to the sky signal and, therefore, the spectral smoothness of the receiver was limited to be somewhat less than an octave.  Additionally, the loss in the hybrid, which was ahead of any amplifiers in the signal path, resulted in loss of sensitivity.  As a consequence of the antenna and receiver design limitations, SARAS~2 was limited to EoR science in the band 115--185~MHz. 

The SARAS~3 receiver, described in detail below, is designed to mitigate these limitations, taking constructive lessons from the SARAS~2 experience. In order to better cancel additives and systematics, the signal from the antenna is directly coupled to a low-noise amplifier through a Dicke switch, which alternately connects the receiver between the antenna and an internal reference load.  Spectral structure arising from multi-path interference within the receiver is reduced by reducing the total electrical length in the front-end by connecting the antenna terminal directly to a miniature RF switch and high-gain modular MMIC amplifier that are followed immediately by an RFoF modulator.  Phase switching to cancel additives in the analog receiver chain and digital samplers is achieved, without loss of sensitivity and without multi-path interference, using a 180-degree hybrid following the optical RFoF demodulator.  Thus SARAS~3 has both Dicke switching as well as phase switching---a double differencing---to more effectively cancel the additives in the measured spectra.  Additionally, the digital receiver for SARAS~3 has been improved to perform a 16384-point FFT instead of the earlier 8192-point, so as to improve detection and rejection of data corrupted by narrow-band RFI.  The FFT implementation architecture was also changed in firmware to be $2\times8$k instead of $8\times2$k, so as to avoid errors at 2k boundaries in the 8k spectrum.

\section{System overview}
\label{sec:sys_overview}

SARAS~3 is a double differencing radiometer, with Dicke-switching between antenna and reference load plus phase switching. The receiver is designed to be split between a front-end unit that is located immediately below the ground plane of a shaped monopole antenna, and connects via optic fiber to a remote back-end some distance away, typically about 150~m.  

This antenna type has been selected so that the antenna terminals are at ground level and there is no significant length of transmission line between the antenna and receiver.  The wideband antenna will almost certainly not have an excellent impedance match across the entire band of operation, which results in internal reflection of system temperature at the antenna terminal.  Avoiding a transmission line between the antenna terminals and receiver avoids delays in the signal path and hence avoids frequency dependent structures arising from multi-path interference.  Second, monopoles are inherently unbalanced at their terminals and hence do not require a balanced-to-unbalanced (balun) transformer to connect to a receiver with a coaxial input terminal.  Third, monopoles with finite conductive ground planes inherently have nulls towards the horizon, which aids in suppression of unwanted terrestrial interference that mostly arrives from low elevations.  Lastly, monopoles may be designed to be electrically short and thus have spectrally smooth reflection and radiation efficiency characteristics, which preserve the spectrally smooth nature of the foregrounds as they couple to antenna temperature.  Short monopoles also have achromatic beams, which prevent mode coupling of foreground spatial features into the frequency spectrum. 

\begin{figure}[htbp]
\begin{center}
\includegraphics[scale=0.48]{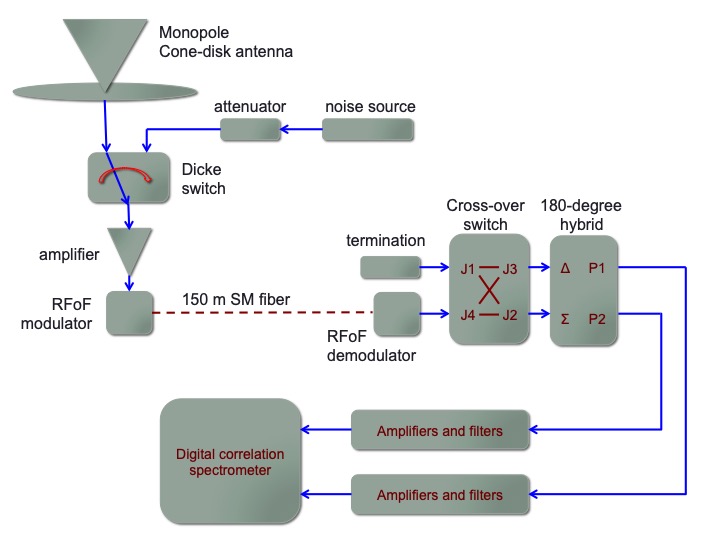}
\caption{Schematic of the SARAS~3 receiver architecture.}
\label{fig:blk_dia}    
\end{center}   
\end{figure}

The signal flow in SARAS~3 is depicted in Fig.~\ref{fig:blk_dia}.  A Dicke switch selects between the antenna and a reference termination.  This reference is a flat-spectrum noise source that is connected to the Dicke switch via attenuations, which ensure that the reference presents a constant impedance independent of whether the calibration noise source is on or off.  Thus the Dicke switch sequentially presents, to the receiver, the antenna temperature, the noise temperature of the reference with noise source off and the noise temperature of the reference with calibration noise source on; we refer to these three switch states and corresponding noise temperatures as OBS, REF and CAL respectively.  Differencing the measurement data with antenna connected with that when the reference is connected provides a differential measurement, which cancels most of the unwanted additives in the receiver chain.  Differencing the measurement data with calibration noise source on with that when it is off provides a bandpass calibration.   

The selected signal voltage---from either the antenna, reference termination or calibration noise source---is amplified and directly intensity modulates a semiconductor laser source, thus providing transmission of the RF signal via single-mode optical fiber from the front-end to the remote part of the receiver electronics.  Conversion from electrical to optical followed by optical to electrical provides excellent galvanic and reverse isolation of the front-end electronics from that at the back end.  This optical isolation is essential since the Dicke switch presents different impedances to the subsequent receiver electronics when it switches between the antenna and reference termination.  Consequently, any standing waves in the receiver chain that reflect from the antenna/reference terminations and result in multi-path interference would change with the position of the Dicke switch, thus resulting in calibration errors.  Standing waves are inevitable owing to impedance mismatches along the signal path, and long transmission lines between the front-end and remote electronics would result in complex and high order spectral structure.  Amplifiers do provide isolation; however, they are limited to the difference between their forward gain and reverse isolation.  Optical modulators/demodulators provide excellent isolation, prevent standing waves existing across the back end and front end electronics.  Thus spectral structure in SARAS~3 is limited to the path length between the antenna terminal, which is coincident with the receiver input, and the optical modulator.  All standing waves downstream of the optical isolation is accurately calibrated out as part of the bandpass calibration and the Dicke switching.
 
Digital control signals that operate the Dicke switch and calibration noise source, located in the front-end electronics, are transmitted from the back-end electronics to the front-end receiver over multi-mode optical fibres. The switch, calibrated noise source, low-noise amplifiers, RF optical transmitters, optical receivers for the digital control signals and the control and monitor circuits of the front-end receiver unit are all powered by a set of Li-Ion battery banks.  The entire front-end receiver is housed in an environmentally protected and electromagnetically shielded enclosure, which is attached below the ground plane of the monopole antenna terminals.  The electromagnetic shielding is vital to prevent coupling and feedback of the signal from the high-gain analog electronics beneath the antenna back to the antenna, which would result in calibration errors.

The remote electronics consists of an analog signal-conditioning unit (ASU) and a digital correlation spectrometer.  The ASU implements phase switching to cancel common-mode additives in this part of the electronics, including the samplers; the ASU and digital spectrometer together operate as a correlation receiver.  To achieve this, the RFoF optical signal is first converted back to electrical in an optical demodulator,  and the electrical signal goes to a cross-over switch.  The other input of the cross over switch has a matched termination.  It may be noted here that the unwanted constant additive from this termination is about 6 mK equivalent antenna temperature, owing to the 44~dB gain between the antenna and the hybrid; this small additive is nevertheless canceled in the Dicke switching.  The cross over switch alternately connects the antenna signal to the sum ($\Sigma$) and difference ($\Delta$) ports of a \ang{180} hybrid.  Thus the hybrid alternately provides a pair of in-phase or a pair of out-of-phase electrical voltage signals at its two output ports, depending on the position of the cross over switch.  The pair of electrical signals are then processed in separate receiver chains: filtered using pairs of low pass and high pass filters to band limit the signal to a 40--230~MHz range and amplified to levels appropriate for sampling by the analog to digital converters (ADCs) of the digital spectrometer.  The gains within the ASU are optimally distributed so that all the amplifiers operate in their linear regimes with intermodulation contributing less than a mK equivalent distortion.  Power levels are set so that the samplers contribute less than about 1\% additional noise power as a result of quantisation, while leaving sufficient head room within the sampler full scale for RFI.

The SARAS~3 digital back-end is an FX correlator based on Virtex 6 field programmable gate array (FPGA) \cite{2020JAI....Girish}.   High-speed 10-bit ADCs sample and digitise the pair of analog signals from the ASU; the ADCs operate at a sampling rate of 500 MS/s to provide an analog bandwidth of 250 MHz.  The samples are first apodised using a four-term Blackman-Nuttall window and then the digital receiver processes blocks of 16k samples in pipelined FFTs.  As the FFT has Hermitian symmetry and the noise equivalent bandwidth of the window is 1.98, the total number of independent channels in the computed spectrum across 250~MHz band is 4096, with a resolution bandwidth of 61~kHz.  The channel data corresponding to the two signals are integrated separately to provide auto-correlation or power spectra; the channel data are also complex multiplied and integrated to provide complex cross-correlation spectra. The real auto- and complex cross-correlation spectra are transferred to a laptop acquisition computer through high speed ethernet and stored as a MIRIAD-format multi-channel visibility data set.   
\begin{table}
\caption{SARAS~3 system states.}
\label{table:sys_states}       
\begin{tabular}{llll}
\hline\noalign{\smallskip}
State & Noise source & Dicke switch & Cross-over switch \\
\noalign{\smallskip}\hline\noalign{\smallskip}
OBS00 & OFF & 0 & 0 \\ 
OBS11 & OFF & 0 & 1 \\ 
CAL00 & OFF & 1 & 0\\
CAL01 & OFF & 1 & 1\\
CAL10 & ON & 1 & 0\\
CAL11 & ON & 1 & 1\\
\noalign{\smallskip}\hline
\end{tabular}
\end{table}

In the complex cross correlation spectrum, the signal amplitude flips sign when the cross over switch alternates.  When the switch feeds the antenna signal to the $\Sigma$ port, the sky signal appears with positive sign and when the switch feeds to the $\Delta$ port the sign of the sky signal flips to negative. However, any common mode additives  and coupling between the pair of channels within the ASU electronics appear in the cross spectra with constant complex correlation; therefore, these unwanted systematics are canceled on differencing the spectra obtained in the two positions of the cross over switch. 

SARAS~3 cycles through six states. The front end cycles between three states, one in which the antenna is connected to the receiver, second where the reference is connected to the receiver, and third when the calibration noise source is on and this noise power flows via the reference attenuation to the receiver.  In each of these states, the cross over switch in the ASU is toggled through its two states.   These six states are given in Table~\ref{table:sys_states}.   In each state, 16 spectral measurements are recorded, each with integration time of 67.11~ms.  Including overheads associated with delays introduced to account for switch settling times and for reading the data into the acquisition computer, the total time for completing a cycle of 6 states is 8.23~s.  

Clocks for the ADCs and the FPGA are derived from a Rubidium atomic standard.  This ensures low clock jitter and sufficient frequency stability so that, given the gradients in filter passbands, frequency drifts within the bandpass calibration cycle time would only result in spectral structure well below 1~mK \cite{2020JAI....Girish}.

The electrical switching signals within the digital spectrometer is potentially a major source of radio frequency interference, if they enter the signal path radiatively via the antenna.  Coupling of self-generated RFI into the analog electronics chains in the ASU is expected to largely cancel in the double differencing.  For this reason, the digital receiver is in a high quality shielded enclosure, with welded walls, absorber lining, and carefully designed enclosure door.  Fans are provided to reject internal heats via RF filtered vents. RF, optical and DC lines are taken through a panel with filtering.  DC power to the digital correlator and ASU is provided from a  battery pack, with RFI blocking filters inserted in these power lines.  Additionally, the antenna is located at a distance greater than 100~m from the digital correlator.   We have made field measurements of the RFI leakage from the digital spectrometer and established that the shielding and separation together suppress the RFI contamination of antenna temperature to below 1~mK \cite{2020JAI....Girish}.  

\subsection{Calibration considerations}
\label{subsec:cal_consi}

The basic calibration of a spectrometer involves bandpass calibration and absolute calibration, which correct for the variation in receiver gain over frequency and provides an absolute temperature scale for the measured spectra. The spectral power available at the antenna terminal is multiplied by the spectral gain of the entire receiver chain of the spectrometer,  impressing a multiplicative band shape on the antenna temperature.  Bandpass calibration removes this instrumental 
bandpass response.  The spectral data recorded by a spectrometer is usually in arbitrary units that is determined by the net electronics gain of the receiver chain and also the arithmetic in the digital signal processing. The process of absolute calibration sets a scale to the counts so that the measurement is converted to be in units of Kelvins of antenna temperature.   Both the bandpass and also absolute gain of the receiver system may be time varying, usually because the physical temperature of the amplifiers and other components may change over time; the SARAS~3 receiver is not actively temperature stabilised.

In SARAS~3, calibration data are derived using the internal calibration noise source, which is connected to the Dicke switch via fixed attenuators.  In the CAL00 and CAL01 states the noise source is off and the system is connected to the 50~$\rm \Omega$ reference termination that is at ambient temperature, which we refer to as $T_{\rm REF}$.  The recorded spectral powers, in arbitrary units, are termed $P_{\rm CAL00}$ for the state when the cross over switch is in position 0, and $P_{\rm CAL01}$ when the cross over switch is in position 1.  Switching on the noise source injects noise of equivalent temperature $T_{\rm CAL}$ into the system and the corresponding spectral powers recorded are denoted by $P_{\rm CAL10}$ for the case when the cross over switch is in position 0 and $P_{\rm CAL11}$ for when the cross over switch is in position 1.  We denote the spectral powers recorded when the Dicke switch connects the antenna to the receiver by $P_{\rm OBS00}$ and $P_{\rm OBS11}$ respectively for the cases where the cross over switch is in position 0 and 1.  We may write
\begin{align}
\label{eq:cal_a}
P_{\rm OBS00} & =  -|G|^2(T_{\rm A} + T_{\rm N}) + P_{\rm cor}, \\
P_{\rm OBS11} & = ~~~|G|^2(T_{\rm A} + T_{\rm N}) + P_{\rm cor}, \\
P_{\rm CAL00} & =  -|G|^2(T_{\rm REF} + T_{\rm N}) + P_{\rm cor}, \\
P_{\rm CAL01} & = ~~~|G|^2(T_{\rm REF} + T_{\rm N}) + P_{\rm cor},\\
P_{\rm CAL10} & =  -|G|^2(T_{\rm CAL} + T_{\rm N}) + P_{\rm cor}, {\rm ~and} \\
P_{\rm CAL11} & = ~~~|G|^2(T_{\rm CAL} + T_{\rm N}) + P_{\rm cor},
\end{align}
where $T_{\rm A}$ is the antenna temperature, $T_{\rm N}$ is the receiver noise added by the analog electronics, $|G|^2$ is the power gain of the receiver system and $P_{\rm cor}$ is the unwanted spectral power added by the samplers of the digital receiver.   

$P_{\rm cor}$ is not expected to change either in power or spectral characteristics between system states; therefore, differencing the powers recorded in the two cross-over switch positions would cancel it:
\begin{align}
\label{eq:cal_b_1}
P_{\rm OBS} & = P_{\rm OBS11} - P_{\rm OBS00},  \nonumber \\
	    & = 2|G|^2(T_{\rm A} + T_{\rm N}). \\
\label{eq:cal_b_2}
P_{\rm REF} & = P_{\rm CAL01} - P_{\rm CAL00},  \nonumber \\
	    & = 2|G|^2(T_{\rm REF} + T_{\rm N}). \\
\label{eq:cal_b_3}
P_{\rm CAL} & = P_{\rm CAL11} - P_{\rm CAL10},  \nonumber \\
	    & = 2|G|^2(T_{\rm CAL} + T_{\rm N}).
\end{align}

A second differencing of the two spectra obtained in Eqs.~\ref{eq:cal_b_3} and \ref{eq:cal_b_2} yields a measure of the system bandpass:
\begin{align}
\label{eq:cal_bp_a}
P_{\rm CAL} - P_{\rm REF} & = 2|G|^2(T_{\rm CAL} + T_{\rm N}) - 2|G|^2(T_{\rm REF} + T_{\rm N}),\nonumber \\
			    & = 2|G|^2(T_{\rm CAL} - T_{\rm REF}).
\end{align}
The factor $(T_{\rm CAL} - T_{\rm REF})$ is the excess noise injected by the calibration noise source when on and we refer to this excess spectral power as $T_{\rm STEP}$. 

Similarly, a second differencing of the two spectra obtained in Eqs.~\ref{eq:cal_b_1} and \ref{eq:cal_b_2} yields a measure of the difference between antenna temperature and that of the internal reference:
\begin{align}
\label{eq:cal_bp_b}
P_{\rm OBS} - P_{\rm REF} & = 2|G|^2(T_{\rm A} + T_{\rm N}) - 2|G|^2(T_{\rm REF} + T_{\rm N}),\nonumber \\
			    & = 2|G|^2(T_{\rm A} - T_{\rm REF}).
\end{align}

Calibrating $(P_{\rm OBS} - P_{\rm REF})$ with $(P_{\rm CAL} - P_{\rm REF} )$ yields the measured temperature 
\begin{align}
\label{eq:cal_bp_TA_TREF}
T_{\rm meas}& = \frac{P_{\rm OBS} - P_{\rm REF}}{P_{\rm CAL} - P_{\rm REF}} T_{\rm STEP} ,\nonumber \\
& =  \frac{T_{\rm A} - T_{\rm REF}}{T_{\rm CAL} - T_{\rm REF}} T_{\rm STEP}.						
\end{align}
The systematic additives have been canceled by the double differencing.  The calibration factor $T_{\rm STEP}$, which determines the absolute scale for
the measured spectrum, is determined by a procedure involving a ``calibration of the internal calibrator" by referencing it to noise power from an external termination whose physical temperature is controlled; this is described in Section~\ref{subsec:abs_cal}. 

Eq.~\ref{eq:cal_bp_TA_TREF} is the spectrum that SARAS~3 measures. 

\section{Measurement equations}
\label{sec:meas_eq}

The calibration of the data recorded in the different switch states is described above in Section~\ref{subsec:cal_consi} and that leads to Eq.~\ref{eq:cal_bp_TA_TREF}, which defines the measured spectrum of the antenna temperature.  In an ideal system design, all the terms except $P_{\rm OBS}$ in the aforementioned equation are frequency independent and hence the calibrated spectrum has a frequency dependence arising purely from the antenna temperature. However, owing to internal reflections of system noise caused by impedance mismatches in the signal path within the receiver chain, the measured spectrum will have frequency dependent terms that may have complex spectral structure. An understanding of these subtle effects is critical in system design as well as selection of suitable components for system realization. In this section, we expand on the description of the measurement equations, taking into account reflections at interfaces. For brevity, only final expressions are given here, details of the derivation may be found in Appendix~\ref{app:measu_equ}.

In the SARAS~3 analog receiver, reflections that occur before the first stage of amplification---in the interconnect and signal path between the antenna and first low-noise amplifier---result in the dominant complex features in the measured spectrum. The important factors to be considered are:
\begin{enumerate}
\item The back and forth reflections of signal power corresponding to the antenna temperature, between the antenna and LNA. 
\item The backward propagation of receiver noise towards the antenna and its back and forth reflections between the LNA and antenna.
\end{enumerate}

The impedance matching of a wideband antenna to a transmission line is relatively more difficult technically compared to matching the transmission line to a wideband low-noise amplifier.  Taking into account first-order reflections of the receiver noise at the antenna, and assuming that the receiver is perfectly matched to the cable connecting it to the antenna, the equation for the calibrated temperature $\rm T_{\rm meas}$ may be written as 
\begin{eqnarray} \nonumber
\label{eq:T1_calib_main}
\rm T_{\rm meas} & = & \rm T_{\rm STEP}\Big[\frac{P_A-P_{REF}} {P_{CAL}-P_{REF} }\Big] + \nonumber \\
&& \rm T_{\rm STEP}\Big[ \frac{P_N}{P_{CAL}-P_{REF}} \times  \Big\{ 2|f||\Gamma_A|cos(\phi_f+\phi_A+\phi) + |f|^2|\Gamma_A|^2 \Big\} \Big],
\end{eqnarray}
where $\rm {P_A}$ and $\rm {P_{REF}}$ are the noise powers corresponding respectively to the antenna temperature $\rm {T_A}$ and the reference termination temperature $\rm {T_{REF}}$.
$\rm {P_N}$ is the receiver noise power, corresponding to the receiver noise temperature  $\rm {T_N}$, that is added to the signal power $\rm {P_A}$ in the low-noise amplifier.
$\rm f=|f|e^{\rm i\phi_f}$ is the fraction of this receiver noise voltage that emerges from the input of the amplifier and back propagates towards the antenna. $\rm \Gamma_A =|\Gamma_A| e^{i\phi_A}$ is the complex reflection coefficient of the antenna: the scattering matrix element S11.  $\phi$ is the phase change arising from 2-way signal propagation in the transmission line connecting the amplifier and the antenna: $ \phi = (4\pi\nu {\it l})/(v_fc)$ where $l$ is the physical length of the line, $c$ is the speed of light in vacuum and $v_f$ is the velocity factor of the transmission line.   All the terms may have a frequency dependence.

The last two terms in Eq. \ref{eq:T1_calib_main} give the spectral additives due to first order reflections of receiver noise at the antenna terminals. If the antenna were perfectly matched to the transmission line connecting to the receiver at all frequencies, $\rm |\Gamma_{\rm A}| = 0$, the contributions from the last two terms would vanish, and this equation reduces to the case in Eq.~\ref{eq:cal_bp_TA_TREF}. However, this condition cannot be satisfied at all frequencies for an antenna, particularly for a wideband antenna, and hence contributions from the reflection terms would inevitably appear in the final spectrum. The cosine term arises from the interference between the forward and reflected noise waves of the receiver, these have a relative phase $(\phi_f+\phi_A+\phi)$.  Although $\phi_f$ and $\phi_A$ would have a variation over the CD/EoR band, it is the length of the transmission line that usually dominates the total change in $(\phi_f+\phi_A+\phi)$ across the band.   Therefore, it is critical to limit the length $l$ and a design goal is to keep the total phase change across the band to within a fraction of  $2\pi$, to maintain the spectral smoothness in $\rm T_{meas}$.  With this aim, in SARAS~3 the length $l$ has been reduced so that the contributions of the phase terms add up to less than $\pi/2$ so that after calibration, the additive contribution of the receiver noise to the measured spectrum is maximally smooth.  Additionally, the antenna is designed to have a maximally smooth reflection coefficient, and the first stage amplifiers are selected to be ultra wideband, so that the characteristics---S11 and also the complex factor $f$---within the CD/EoR band are maximally smooth.  Separately, reduction in the magnitude of the reflected receiver noise is achieved by using a Low Noise Amplifier (LNA), with low noise temperature, as the first amplifier in the receiver.  

We next relax the assumption that the LNA is matched to the transmission line. This would introduce additional components in the measured spectrum, as both the signal from the antenna and the backward-propagating receiver noise from the low-noise amplifier undergo multiple reflections between the antenna and LNA input.   We may view the transmission line connecting the antenna to the amplifier as a leaky resonant cavity that supports various resonant modes, with the power coupled to the LNA and antenna as leakages. Taking into account the series of higher order reflections, the expression for the measured spectrum may be written in the form
\begin{align}
\label{eq:Tinf_calib_main}
\rm T_{\rm meas} & = \rm T_{\rm STEP}\Big\{ \rm  \frac{P_A [ \sum_{k=0}^{+\infty}~|\Gamma_N|^k |\Gamma_A|^k \sum_{l=0}^{k} cos\{(2l-k)(\phi_N+\phi_A+\phi)\}]-P_{REF}} {P_{CAL}-P_{REF} } \\ \nonumber
& + \rm \frac{P_N}{P_{CAL}-P_{REF}} \times \Big[\sum_{k=0}^{+\infty} (2|f||\Gamma_A|^{(k+1)}|\Gamma_N|^k cos\{\phi_f +(k+1)(\phi_A+\phi) +k\phi_N\}) \\ \nonumber
&~~~~~~~~~~~~~~~~~~~~~~~~~~~ +  \rm |f|^2|\Gamma_A|^2\sum_{k=0}^{+\infty}~|\Gamma_N|^k |\Gamma_A|^k \sum_{l=0}^{k} cos\{(2l-k)(\phi_N+\phi_A+\phi)\} \Big]\Big\}.
\end{align}

The above Eq.~\ref{eq:Tinf_calib_main} is a detailed expression for the measured spectrum.  First, it is clear that better impedance matching between the antenna, the transmission line that follows, and the low-noise amplifier at the end of the transmission line, are critical.  Improving either or both these will substantially reduce the amplitude of successive reflections and hence reduce the amplitude and complexity of unwanted structures in the measured spectrum.  

In principle, it is possible for all the quantities in Eq.~\ref{eq:Tinf_calib_main} to be measured---both via laboratory calibrations and field measurements---and the measured sky spectrum may be then corrected for the multiple unwanted terms to derive an estimate of the antenna temperature.  To examine the validity of the derived model, we provide below in Sec.~\ref{subsec:meas_equ_fit_data} the results of fitting Eq.~\ref{eq:Tinf_calib_main} to laboratory measurements made with the antenna replaced with precision electrical open and short terminations.  To model the measurement equation to the accuracy necessary for detection of CD/EoR signals, such an approach requires precise laboratory and in-situ measurements of several quantities, necessitating complexity in system design to allow for the calibrations.  

As mentioned above, the SARAS approach has been to design the receiver to avoid precise modelling of the terms in the detailed measurement equation.  The SARAS~3 design strategy has taken an alternate path recognising that the complexity of the unwanted spectral structures may be substantially reduced by reducing the path length over which the system temperature components suffer multiple internal reflections.  As the total differential path increases with multiple reflections, the interference between the signals with larger differential paths results in higher order structure in the measured spectrum.

It may be noted here that the only relevant paths are those which involve the antenna terminals at one end, since it is only those terminals that are Dicke switched over to the reference port for calibration.  Internal reflections between any pair of impedance mismatches downstream of the low-noise amplifier would result in bandpass structure, which would be calibrated out.  In a radiometer receiver chain where the antenna is followed by successive stages of amplification, each amplifier provides an isolation for multi-path propagation that is limited to the difference between the reverse loss and forward gain.  Thus the length $l$ over which uncalibrated multi-path reflections may occur is not limited to just the length between the antenna and first low-noise amplifier, but may be the effective length between the antenna and many stages of amplification, depending on the effective isolation provided by successive components.  Therefore, the SARAS~3 design has introduced optical isolation immediately following the first amplification module, by introducing an optical modulator, fiber optic transmission line and a demodulator.  Thus for SARAS~3 the only relevant path is that between the antenna and the optical isolator; this path has been made much smaller than the wavelengths of operation.

Finally, we qualify our receiver by evaluating the receiver performance by replacing the antenna with precision loads as well a load with frequency-dependent reflection coefficient similar to that of the CD/EoR antenna and examining for the level of confusion between spectral structures arising from uncalibrated internal systematics and plausible CD/EoR signals. Details of these tests are given below in Section~\ref{sec:lab}.

\section{Implementation of the SARAS~3 receiver}
\label{sec:implementation}

The SARAS~3 receiver is implemented in two sections: an antenna base electronics and a remote station electronics unit that is located about 150~m away.  The radio frequency (RF) signal from the antenna base electronics travels as RF over fiber (RFoF), intensity modulated on a laser carrier and in a single-mode fiber, to the remote station analog electronics units for further signal processing.   The entire analog receiver chain is designed to operate in the band 40 to 230 MHz that is defined by high and low pass filters; however, when operated in sites where FM might be present with intolerable strength, the low pass filters may be replaced with units that cut the band above 87.5~MHz for CD signal detection, or the high pass filters may be replaced with units that cut the band below 110~MHz for EoR signal detection.  Control signals that time the switching of states in the analog sections are generated in a control and acquisition computer that is part of the digital receiver, and sent to both analog units via multi-mode optical fiber.  The digital receiver is co-located with the remote analog electronics unit about 150~m from the antenna and the electronics at its base.

\begin{figure}[htbp]
\begin{center}
\includegraphics[scale=0.48]{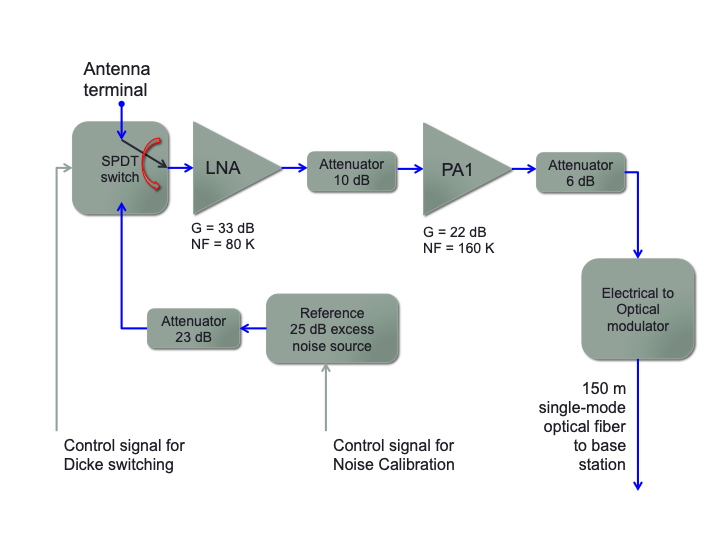}
\caption{SARAS~3 analog electronics at the base of the antenna.}
\label{fig:base_electronics}    
\end{center}   
\end{figure}

The antenna base electronics unit has the components in Fig.~\ref{fig:base_electronics} in the signal path.  An electro-mechanical surface-mount RF switch with an insertion loss less than 0.1~dB and isolation greater than 60~dB selects between the antenna temperature and that from a reference ambient temperature termination.  This reference is implemented as a matched 50-${\rm \Omega}$ 23~dB attenuator connected to a flat-spectrum noise source of 25~dB excess noise.  Calibration signal is input to the radiometers when this noise source is on; the radiometer sees a reference termination when off.  Interconnects between the switch and terminals of the antenna and reference use adaptor bullets, without cables, to minimize path lengths.  When the control to the switch is off, the analog receiver chain is disconnected from the antenna thereby providing protection from static.  Protection during observing is provided with Schottky diodes and PIN diodes to prevent environmental electrostatic discharge and high power RFI damaging the sensitive low-noise amplifier. These devices also provide protection from human body static when the antenna is manually mounted on the receiver.

The first active device is a MMIC based low-noise amplifier with gain of 33~dB, followed by a second amplifier providing an additional gain of 22~dB.  A chip attenuation of value 10~dB is placed in the path between the two for improved impedance matching  and isolation.  The first amplifier has noise temperature of about 80~K and the second 160~K.  The switch, protection devices, amplifiers, attenuations, along with a final attenuation of 6~dB following the second stage of amplification, are all accommodated in a single custom-made printed circuit board (PCB) designed in-house and mounted in a compact aluminum enclosure.  A modular electrical to optical RFoF modulator follows that is based on a distributed feedback (DFB) laser and provides intensity modulated 1310~nm laser light on a single-mode fiber.  

No filtering is done in the antenna base electronics!  This difficult decision was made based on the critical requirement of keeping the signal path between the antenna terminals and optical isolator as small as possible. Thus the signal bandwidth at this stage is determined by the antenna bandwidth and that of the pair of amplifiers.

\begin{figure}[http]
\begin{center}
\includegraphics[scale=0.48]{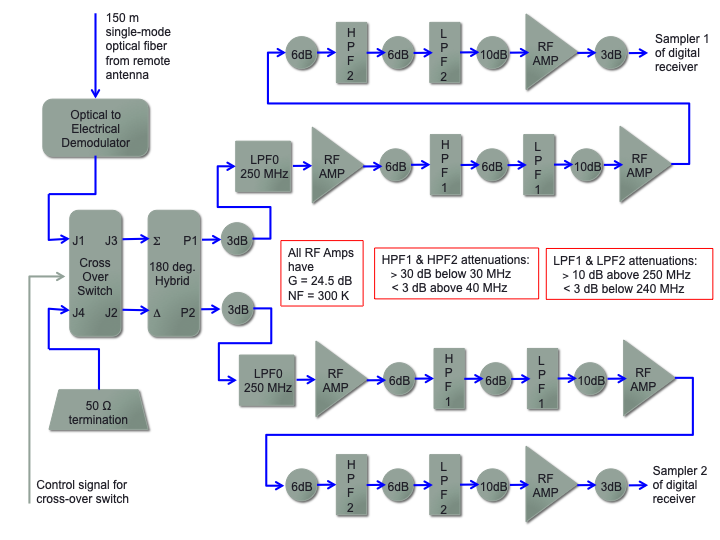}
\caption{SARAS~3 analog electronics at the remote station 150~m from the antenna.}
\label{fig:remote_electronics}    
\end{center}   
\end{figure}

The remote station analog electronics unit has the components shown in Fig.~\ref{fig:remote_electronics}.  The key block in this unit is a cross-over switch, implemented as a high-reliability electro-mechanical coaxial switch, that is connected to the sum and difference ports of a \ang{180} hybrid.  The RFoF signal is first converted back to RF in an optical to electrical demodulator and this signal forms one input to the block.  The ambient-temperature noise power of a 50-${\rm \Omega}$ matched termination forms the second input.  The cross-over switch has an isolation better than 80~dB in the receiver operating band.  The two outputs P1 and P2 are the sum and difference of these two inputs, and toggling the cross-over switch causes these outputs to swap.  The block implements phase switching and provides a pair of signal outputs for a correlation receiver.  

The pair of signals from the block go through identical receiver chains.  The signals are first low-pass filtered to about 250~MHz with coaxial Butterworth filters.  Then follows three stages of amplification with sharper low and high pass filters in between, to limit the 6-dB bandwidth to 40--240~MHz and have 60~dB attenuation below 30~MHz and above 260~MHz.  These filters are designed and developed in-house using discrete inductors and chip capacitors realizing 9th-order Elliptic filter approximations.  The amplifiers in the remote station receiver chains, where power levels are higher compared to the antenna base electronics chain, are built using modules with 24~dB gain and high 1~dBc of +24~dBm power level, which leaves a headroom exceeding 40~dB for RFI within the band.  The band-limited signals are available for the digital receiver.

All of the antenna base electronics are powered by a pair of Li-Ion battery packs of 18~V/20~AH rating; only one battery pack is used at any time and the power source may be switched remotely.  Together they are capable of operating the receiver for eight nights of observing before the receiver needs to be accessed for charging the battery packs.  Linear regulators with low dropout ratings are used to supply the different voltages required; the voltages are supplied to the different modules via shielded coaxial cables to obtain about 40~dB of isolation to radiated couplings within the receiver enclosure. The power supply lines are provided with good filtering at both the source and destination ends to ensure that these lines are not a source of feedbacks for RF power.  

\begin{figure}[!ht]
\begin{minipage}[!ht]{0.42\linewidth}
\includegraphics[width=\linewidth]{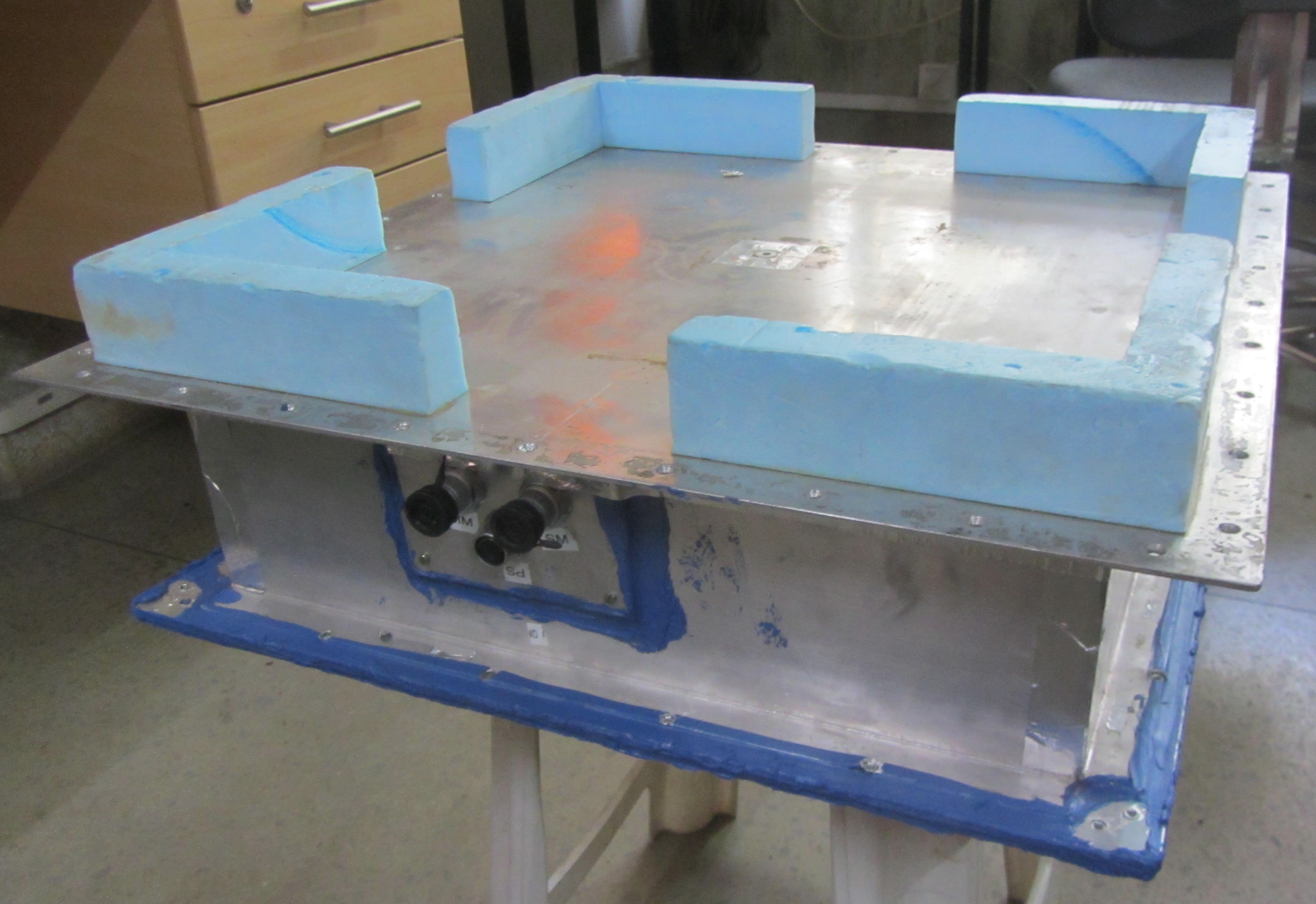}
\end{minipage}
\begin{minipage}[!ht]{0.36\linewidth}
\includegraphics[width=\linewidth]{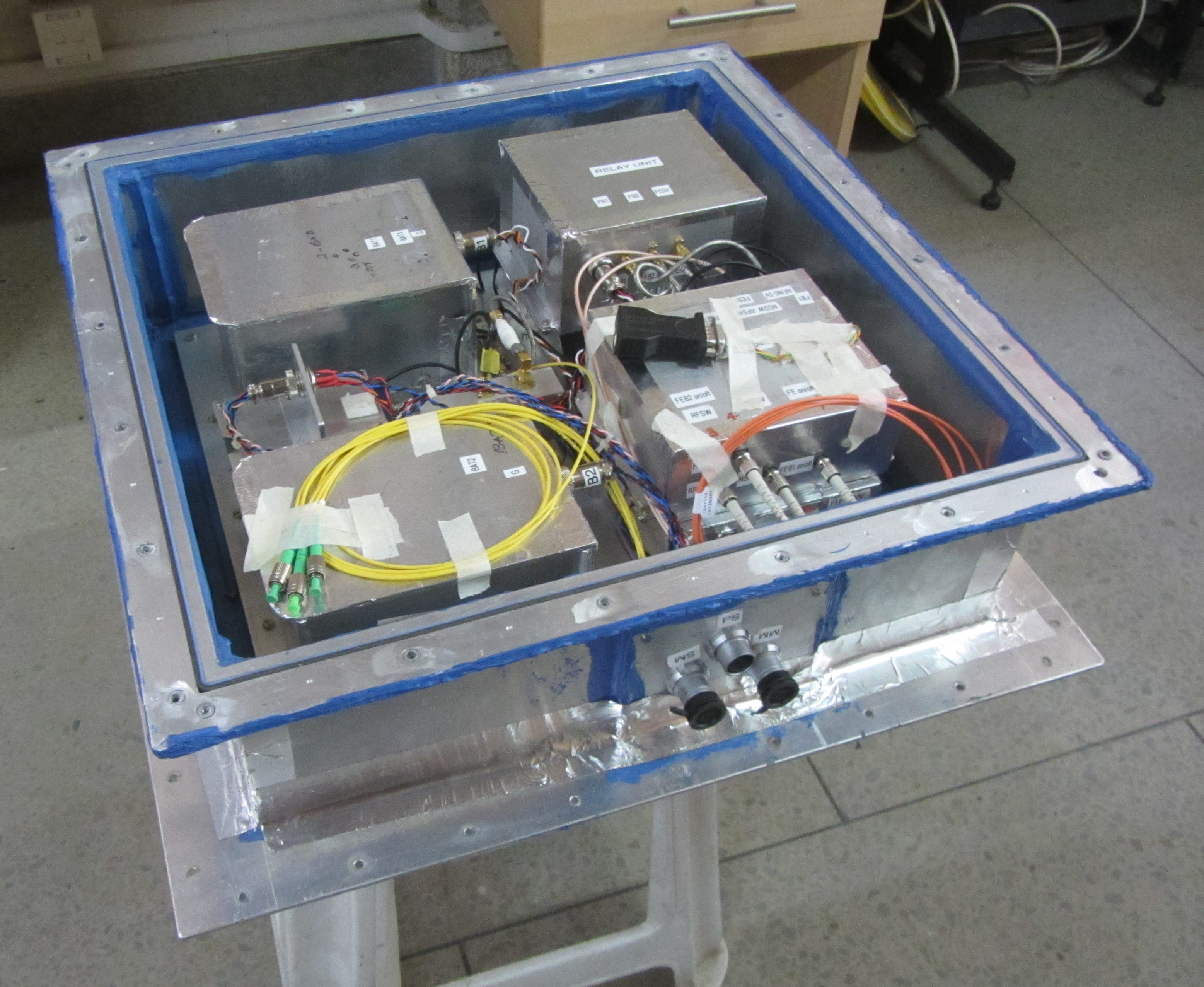}
\end{minipage}
\begin{minipage}[!ht]{0.21\linewidth}
\includegraphics[width=\linewidth, angle=0]{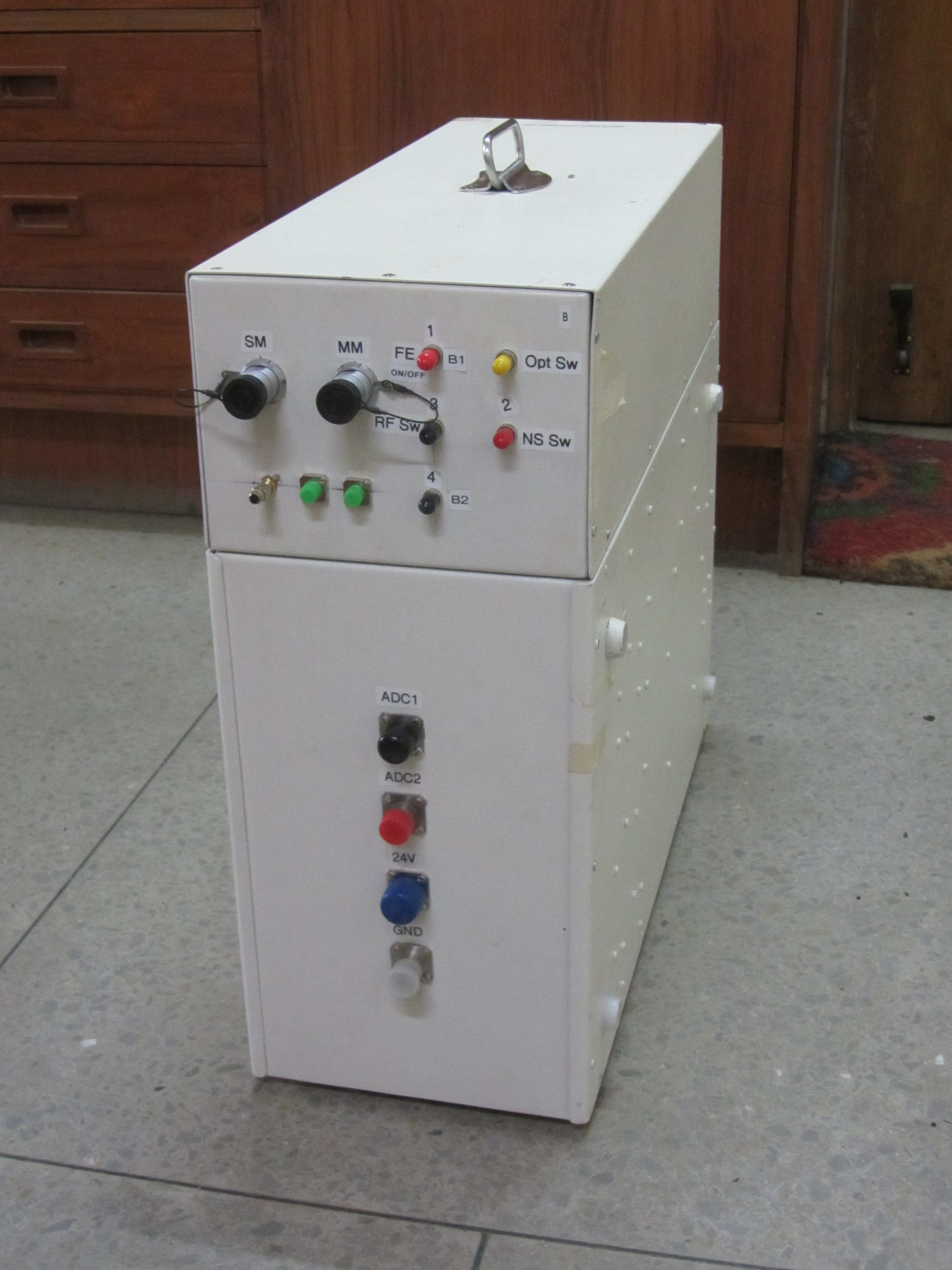}
\end{minipage}
\caption{The analog electronics enclosures beneath the antenna and at the remote base station.  In the panel on the left and middle are shown the enclosure at the antenna base; the left panel shows its view from the top and in the middle panel is shown the view from below with the bottom cover removed.   The panel on the right shows the enclosure that houses the analog electronics at the remote station.}       
\label{fig:receiver_images}
\end{figure}

The antenna base electronics is housed in a square aluminum welded box. The top face of the box is also part of the ground plane of the antenna and has a UHF to SMA adaptor at its center for the monopole terminal of the antenna.  This top face has mounting holes along flanges at its edges where extensions to the ground plane may be fastened to extend the ground plane to any desired area around the monopole.  Styrofoam blocks are glued on to guide the placement of the monopole to the connector and prevent lateral movement.  This top view of the antenna base electronics box is shown in the first panel in Fig.~\ref{fig:receiver_images}.

A photograph of the antenna base electronics enclosure, flipped over and with the cover removed, is in the middle panel of Fig.~\ref{fig:receiver_images}.  The electronics modules are mounted on a plate fixed to the top panel of the enclosure, and a flat cover is bolted on below.  All components are mounted in separate aluminum chassis even within the enclosure and these are sealed with screw-on lids and covered along mated edges with Aluminum tape.  Electromagnetic sealing and water proofing of this enclosure is provided by having 'O'-ring groves machined on the flange, in which gaskets are placed that provide both these protections.  Additionally, silicone waterproofing sealant is applied all along edges where the lid mates with the box.  Connectors for the single-mode and multi-mode fiber cables, plus an additional connector that provides access to the batteries within the enclosure for charging, are in a plate mounted on one side of the enclosure.

Within the antenna base enclosure is a temperature logger that records the physical temperature of the reference termination.  This provides the noise temperature of the load that is Dicke switched in place of the antenna.  The temperature of the top ground plate of the receiver enclosure is also recorded.  The logger is accessed only after an observing campaign is completed and the sealed enclosure is opened.

The remote station analog electronics is in an enclosure shown in the last panel in Fig.~\ref{fig:receiver_images}.  The optical components are in a separate segment at the top, and the electrical modules are mounted on the inner side walls of the enclosure.  The walls are hinged and may be opened on either side to access the receiver chains that form the two arms of the correlation receiver.  This enclosure receives 24~V d.c. power via a pair of coaxial cables from a battery box, and control signals via four multi-mode fibers from the digital receiver.  Single-mode and multi-mode fiber cable connectors are provided for the 150-m outdoor fiber-optic cables that run from this remote station to the antenna base electronics enclosure.  A pair of panel mounted coaxial connectors are provided for cabling the processed RF power to the digital correlation spectrometer.

The RFoF link is provided by a rugged waterproof fiber cable with IP65 rating and single-mode APC connectors.  The RFoF link has an overall gain of 5~dB and a noise figure of 19~dB, or 23,000~K.  The pre-amplifier of the optical modulator thus contributes about 3~K system temperature to the overall noise budget.  It is the RFoF link that limits the headroom for RFI to about 30~dB above sky noise.

\section{Sensitivity of the SARAS~3 receiver}
\label{sec:sensit}

The sensitivity of a receiver depends on the thermal noise that the calibrated spectra have, as well as any residual systematic errors. By careful design, SARAS~3 receiver and antenna have maximally smooth responses and hence any residual systematic is expected to be less than a mK; this is demonstrated below in Section~\ref{sec:lab}. Therefore, the sensitivity of the system depends predominantly on the random measurement noise in the spectra recorded in each of the six states through which the system cycles, and the mechanics of calibration that combines these recordings to compute calibrated spectra for the antenna temperature. 

In the SARAS~3 receiver, spectral powers measured in each of the six switched states would have different associated noise. For measurement data recorded in each of these states, associated uncertainties are computed and stored as metadata in the pre-processing stages of the software pipeline. The sensitivity of the system is then estimated by propagating these uncertainties through the calibration equation given in Eq.~\ref{eq:cal_bp_TA_TREF}, which is reproduced here for reference:
\begin{align}
\label{eq:cal_Tmeas_sens}
T_{\rm meas} = \frac{P_{\rm OBS} - P_{\rm REF}}{P_{\rm CAL} - P_{\rm REF}} T_{\rm STEP}.						
\end{align}
All terms in the above equation are functions of frequency and the calibration is computed separately for each spectral channel.  

$P_{\rm OBS}, P_{\rm REF}$ and $P_{\rm CAL}$ individually represent differences between the spectral powers measured in a pair of states, where the phase of the antenna signals is switched relative to internal additives in the arms of the correlation receiver.  We denote the rms noise associated with each of them as $\Delta P_{\rm OBS}, \Delta P_{\rm REF}$ and $\Delta P_{\rm CAL}$ respectively. For small perturbations, the noise in $T_{\rm meas}$ can then be approximately expressed as a combination of the rms noise in the three power measurements:
\begin{align}
\label{eq:del_Tmeas}
\Delta T_{\rm meas} = T_{\rm STEP} \sqrt {\Big (\frac{\partial T_{\rm meas}}{\partial P_{\rm OBS}}\Delta P_{\rm OBS}\Big)^2 + \Big (\frac{\partial T_{\rm meas}}{\partial P_{\rm REF}} \Delta P_{\rm REF}\Big)^2 + \Big (\frac{\partial T_{\rm meas}}{\partial P_{\rm CAL}} \Delta P_{\rm CAL} \Big)^2}.
\end{align}
Evaluating the partial derivatives yields
\begin{align}
\label{eq:del_Tmeas_eq}
\Delta T_{\rm meas} &=& T_{\rm STEP} \Big[ \Big\{\frac{\Delta P_{\rm OBS}}{P_{\rm CAL} - P_{\rm REF}} \Big\}^2 + 
\Big\{ \frac{(P_{\rm CAL} - P_{\rm OBS}) \Delta P_{\rm REF} }{(P_{\rm CAL} - P_{\rm REF})^2} \Big \}^2 + \nonumber \\
& & \Big \{\frac{(P_{\rm OBS} - P_{\rm REF}) \Delta P_{\rm CAL} }{(P_{\rm CAL} - P_{\rm REF})^2} \Big \}^2 \Big]^{1/2}.
\end{align}
In the above equation, each of the powers $P_{\rm OBS}$, $P_{\rm REF}$ and $P_{\rm CAL}$ may be expressed in terms of the system temperatures ($T_{\rm OBS}+T_{\rm N}$), ($T_{\rm REF}+T_{\rm N})$ and $(T_{\rm CAL}+T_{\rm N})$ in the respective states using the general form $P = 2 |G|^2T$, where $G$ is the system voltage gain. The rms noise in these measured spectral powers are related to the corresponding system temperatures by the radiometer equation:
\begin{align}
\label{eq:Rad_eq}
\Delta P =  \frac{2|G|^2 T}{\sqrt{B\tau}},
\end{align}
where $\tau$ is the integration time and $B$ is the noise-equivalent bandwidth of the spectral channels. It may be noted here that the gain term $G$ cancels when the substitutions are made and therefore precise information regarding the system gain as a function of frequency is not required for estimating the sensitivity.

We may now estimate the rms noise in calibrated spectra.  As stated above, the receiver system is designed to operate over frequencies 40--230~MHz, and intended to be used with scaled conical monopole antennas that operate in octave bands and cover the range in staggered bands.  We compute here the sensitivity to Cosmic Dawn and Reionization signals when fitted with the floating cone-disc antenna \cite{2020ITAP....Raghu} covering an octave band from 43.75 to 87.5~MHz.   In this band the sky brightness is a maximum and hence sensitivity is the lowest; therefore, bands at higher frequencies will have greater sensitivities and lower rms noise for same integration times.  

We first estimate the noise at 70~MHz.  Away from the Galactic plane, the sky temperature at this frequency is about 2000~K.  The receiver noise is about 80~K and we assume that the reflection efficiency of the antenna is 70\% and radiation efficiency is 50\%.   The system temperature when the switch is in OBS state and the receiver is connected to the antenna would be about 930~K, with 150~K contribution coming from the resistive loss in the environment of the antenna.  The calibration step $T_{\rm STEP}$ of the noise injection corresponds to a temperature of 630~K when referred to the antenna terminals.  In the reference (REF) and calibration (CAL) states, the system temperatures would be about 380~K and 1010~K respectively.  The SARAS correlator provides spectra with a spectral resolution  of 61~kHz, corresponding to 4096 spectral channels over 250~MHz \cite{2020JAI....Girish}.   For spectra with this  resolution, and integration time of 2.7 seconds in each of OBS, REF and CAL states, using Eqs. \ref{eq:del_Tmeas_eq} and \ref{eq:Rad_eq}, the sensitivity is 4.2~K per channel, per spectra. This is the rms noise in a total observing time of $2.7 \times 3 = 8.2$~s.  For an observation session spanning eight hours, there would be about 3400 such spectra, giving a sensitivity of about 73~mK per channel.  If this averaged spectrum were smoothed to a noise-equivalent bandwidth of 4~MHz, the rms measurement noise would be 9~mK.  Referred to the sky, by accounting for the total efficiency, the rms measurement noise in the estimate of sky brightness temperature would be 25.7~mK.

\begin{figure}[http]
\begin{center}
\includegraphics[scale=0.3]{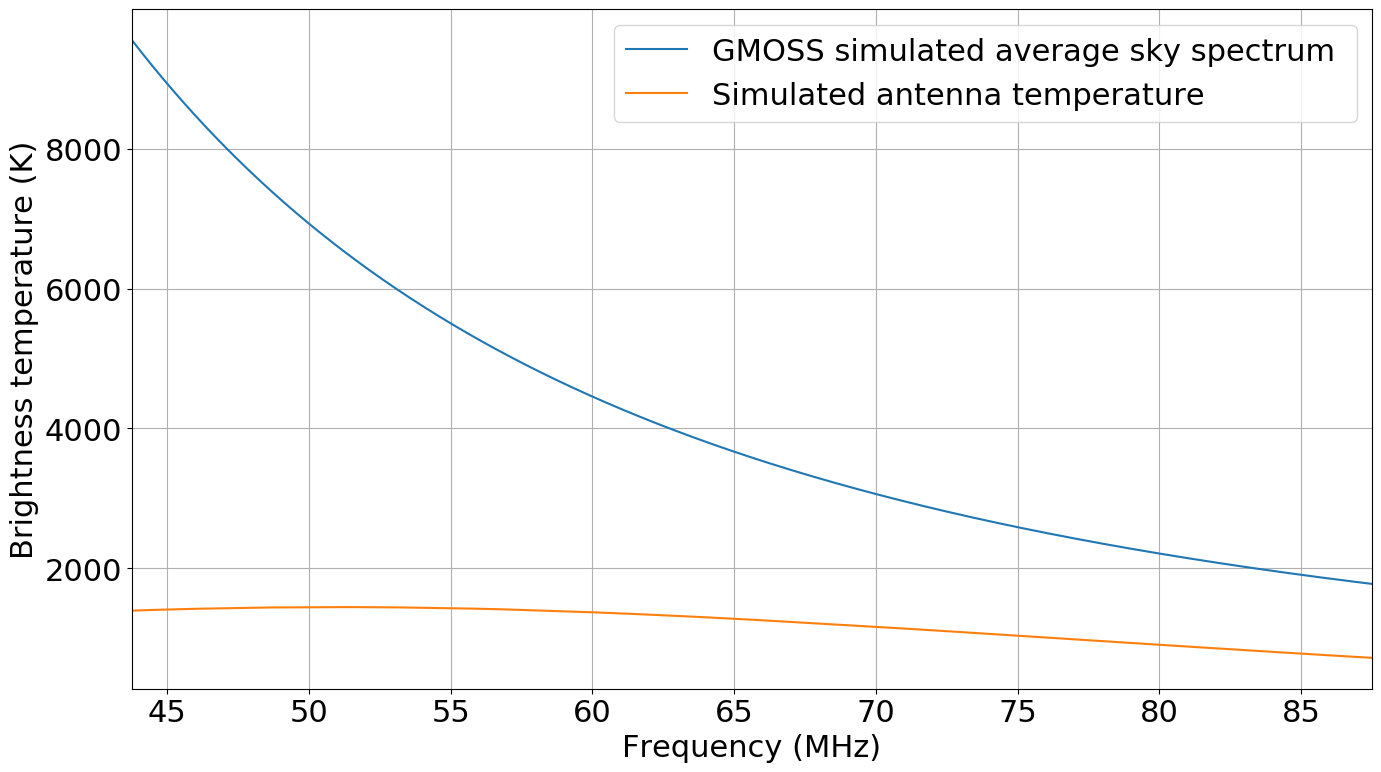}
\caption{Expected antenna temperature for an observation with the SARAS~3 antenna, at latitude $+14^{\circ}$ and over local sidereal time (LST) from 10 to 18~hr.  Also shown is the expected average sky spectrum.}
\label{fig:simulated_Ta}    
\end{center}   
\end{figure}

The above computation may be extended to estimate the expected distribution of rms noise across the octave band, using measured antenna efficiencies and sky models.  We use measured reflection and radiation efficiencies given in Raghunathan et al. \cite{2020ITAP....Raghu} for the SARAS~3 antenna, and use GMOSS \cite{2017AJ....153...26S} model foreground.  We assume that the observing is at latitude $+14^{\circ}$ and over local sidereal time (LST) from 10 to 18~hr that includes a transit of the Galactic plane across the antenna beam. The expected average sky spectrum along with the expected calibrated spectrum of the antenna temperature, over an octave band from 43.75 to 87.5~MHz, is given in Fig.~\ref{fig:simulated_Ta}. 

\begin{figure}[http]
\begin{center}
\includegraphics[scale=0.3]{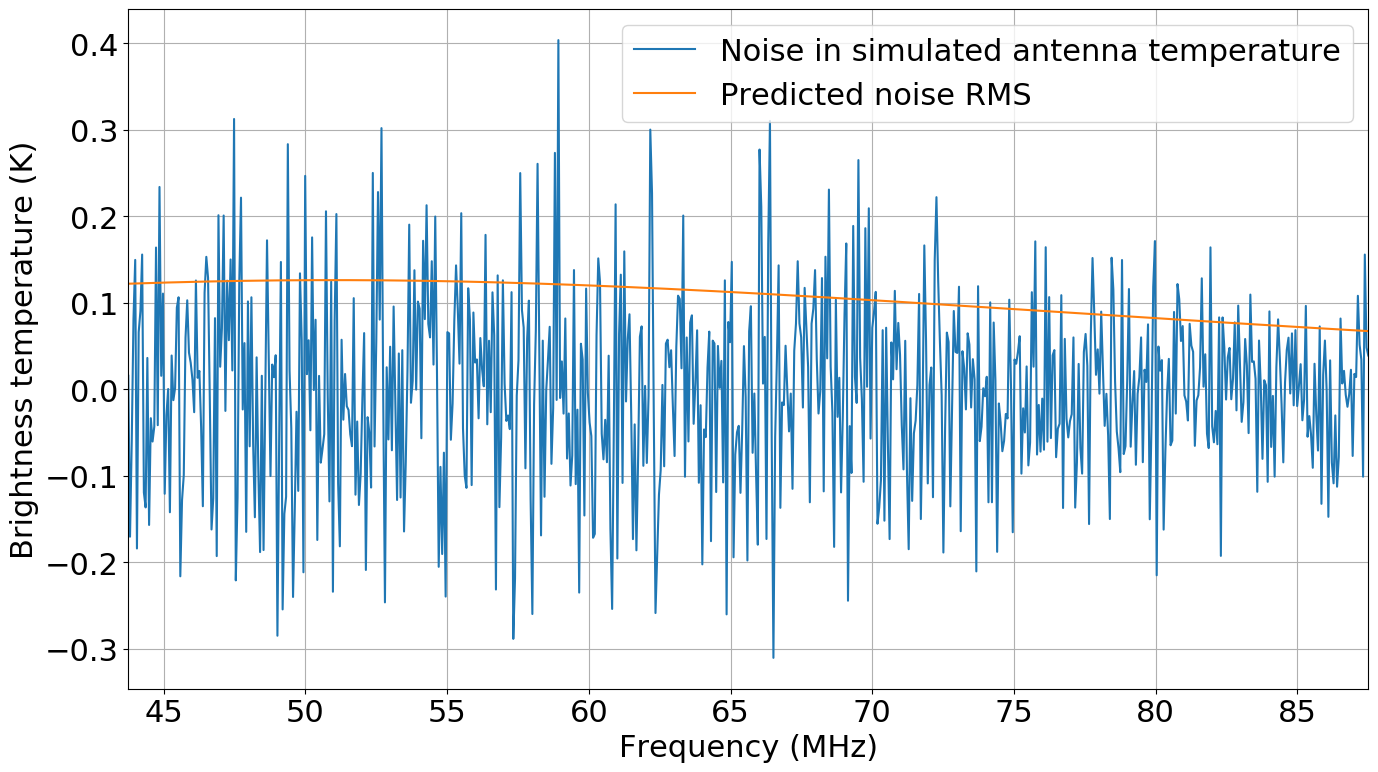}
\caption{An example of the noise component that might be present in an 8~hr mock observation.  The rms noise expected across the band is also shown in orange.}
\label{fig:simulated_Ta_noise}    
\end{center}   
\end{figure}

The OBS data acquired with receiver connected to sky are then calibrated using Eq.~\ref{eq:cal_Tmeas_sens} with CAL and REF spectra acquired with receiver switched to reference and with calibration noise on, respectively.  The rms noise distribution in these spectra are computed using Eqs.~\ref{eq:del_Tmeas_eq} and \ref{eq:Rad_eq} and shown in Fig.~\ref{fig:simulated_Ta_noise}.  It may be noted that, as expected, the rms noise reduces across the band towards higher frequencies since the system temperature is sky dominated and the sky temperature reduces with frequency.

\begin{figure}[http]
\begin{center}
\includegraphics[scale=0.26]{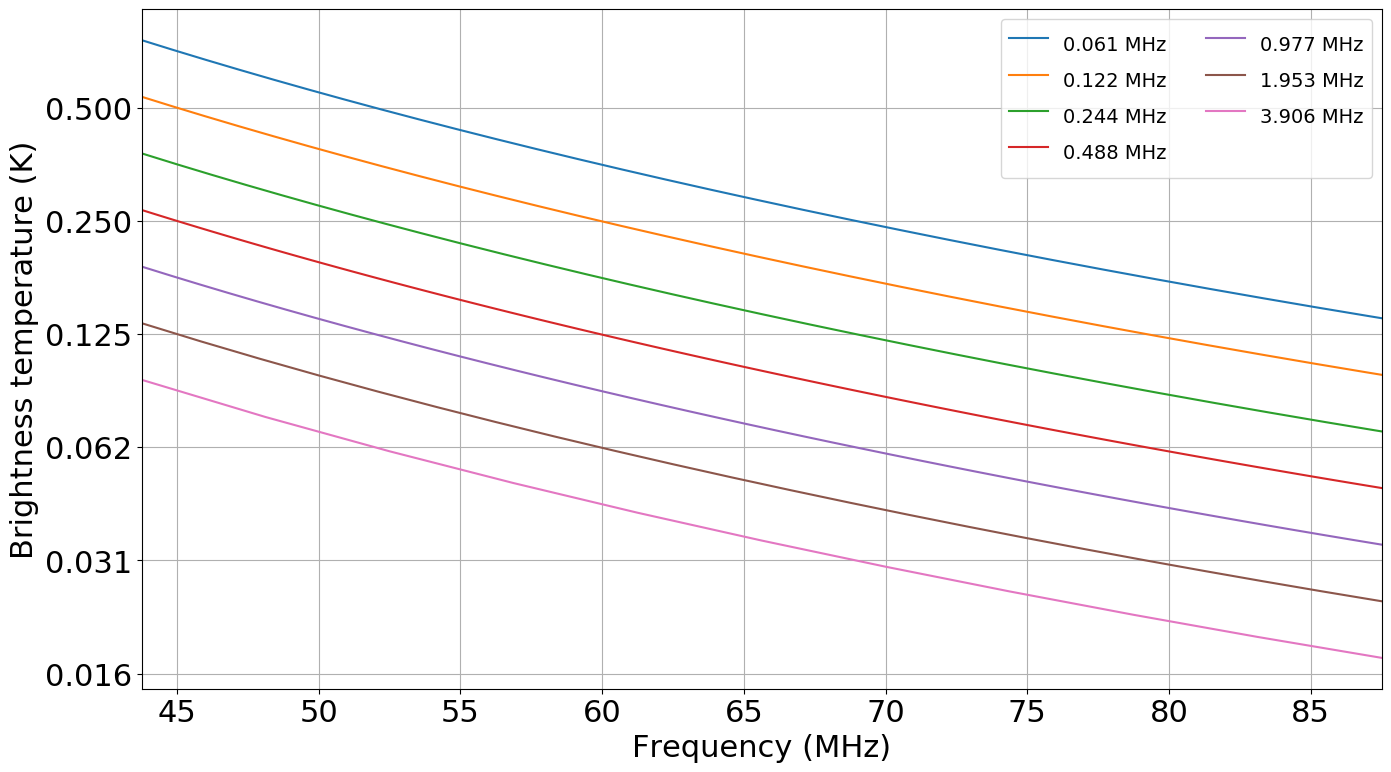}
\caption{The distribution in rms noise for smoothing to different noise equivalent widths, for the 8-hr mock observation.  It may be noted that the rms noise estimate has been referred to the sky domain by correcting for the total efficiency of the radiometer; therefore, the rms noise represents 1-$\sigma$ uncertainty in measurement of sky brightness temperature.}
\label{fig:simulated_Ta_noise_smoothing}    
\end{center}   
\end{figure}

The native spectral resolution of the SARAS digital spectrometer has been designed to be 61~kHz, much finer than the characteristic scale of expected global 21-cm signal, so that any man-made narrow band radio frequency interference (RFI) may be identified and rejected.   Since the global 21-cm signal is expected to have broad spectral shapes, we may smooth the spectra---following rejection of data corrupted by RFI---to a resolution much poorer than the native spectral resolution of 61~kHz of the digital spectrometer, to increase sensitivity without significant loss of signal.  The effect of smoothing on the rms of the measurement noise is shown in Fig.~\ref{fig:simulated_Ta_noise_smoothing}, where the distribution of the rms noise across the band is plotted for smoothing to a range of noise equivalent widths.  If the measurement noise is Gaussian random, the incremental gain in sensitivity is roughly proportional to the square root of number of independent points across which smoothing is performed. 

\section{Laboratory tests}
\label{sec:lab}

\subsection{Absolute calibration}
\label{subsec:abs_cal}

Absolute calibration of the measured data is the process by which the acquired data in some arbitrary counts are converted to units of kelvin using a scaling factor. In the calibration equation given in Eq.~\ref{eq:cal_bp_TA_TREF}, $T_{\rm STEP}$ is a scaling factor that sets the overall temperature scale by virtue of its multiplication with the dimensionless ratio of powers.  It may be noted here that this calibration of the measurement data with $T_{\rm STEP}$ sets the data to be antenna temperature in kelvin scale at a reference point that is the input to the switch, which is same as the terminals of the antenna.  The value of the scaling factor $T_{\rm STEP}$ is determined by the calibrator assembly formed by the noise source and the attenuators that follow it; the receiver is switched between the antenna and this calibrator assembly. Though it is possible to compute $T_{\rm STEP}$ using the published excess noise ratio (ENR) of the noise source and the values of the attenuators between the noise source and switch; for improved accuracy, a laboratory measurement of $T_{\rm STEP}$ is required.  

Laboratory measurement of $T_{\rm STEP}$ is carried out by replacing the antenna with a source of RF noise whose spectral power is known. For this measurement, we replaced the antenna with a precision $\rm 50~\Omega$ termination, and assume that the noise power per unit bandwidth from this matched impedance is given by $k_{B}T$, where $k_{B}$ is the Boltzmann constant and $T$ is the physical temperature of the termination.  Accurate temperature probes are firmly attached to this termination and to the internal reference formed by the attenuators; thermal insulation is provided so that the thermal resistance between the probes and termination/reference attenuator is significantly lower compared to that between the probes and environment. 

Raw (uncalibrated) spectral data are recorded with the receiver cycled sequentially between the termination, reference with noise source off, and reference with noise source on.  Physical temperatures of both the termination and the reference are logged.  For this experimental setup, when the recorded data are calibrated using Eq.~\ref{eq:cal_bp_TA_TREF}, the calibrated spectrum is ideally expected to be the difference between the physical temperatures of the termination and the reference ports, if the value of $T_{\rm STEP}$ is accurate.  The expectation is that $T_{\rm meas} = T_{\rm 50} - T_{\rm REF}$, where $T_{\rm 50}$ and $T_{\rm REF}$ are the physical temperatures of the $\rm 50~\Omega$ termination and reference port respectively.

Despite the experimental setup in which thermal resistance is added between the environment and termination/reference, and care is taken to bond the temperature probes to the termination/reference attenuation, in practice a finite and significant offset was unpreventable between the temperatures logged by the probes and true noise temperatures of the termination/attenuation.   The offset errors may be written in the form: $T_{\rm 50} = T_{\rm 50,m} + T_{ \rm os_1}$ and $T_{\rm REF} = T_{\rm REF,m} + T_{\rm os_2} $, where $T_{\rm 50,m}$ and $T_{\rm REF,m}$ are the temperatures of the $\rm 50~\Omega$ termination and reference as measured by the probes, and $T_{\rm os_1}$ and $T_{\rm os_2}$ denote the offsets in temperatures, which may be positive or negative. Substituting these relations including error terms into Eq.~\ref{eq:cal_bp_TA_TREF}, we obtain:
\begin{align}
\label{eq:T_diff_abscal}
T_{\rm 50,m} - T_{\rm REF,m} = T_{\rm STEP} \frac{P_{\rm OBS} - P_{\rm REF}} {P_{\rm CAL} - P_{\rm REF}} + T_{\rm os},
\end{align}
where $T_{\rm os} = T_{\rm os_2}-T_{\rm os_1}$. 

\begin{figure}[htbp]
\begin{center}
\includegraphics[scale=0.3]{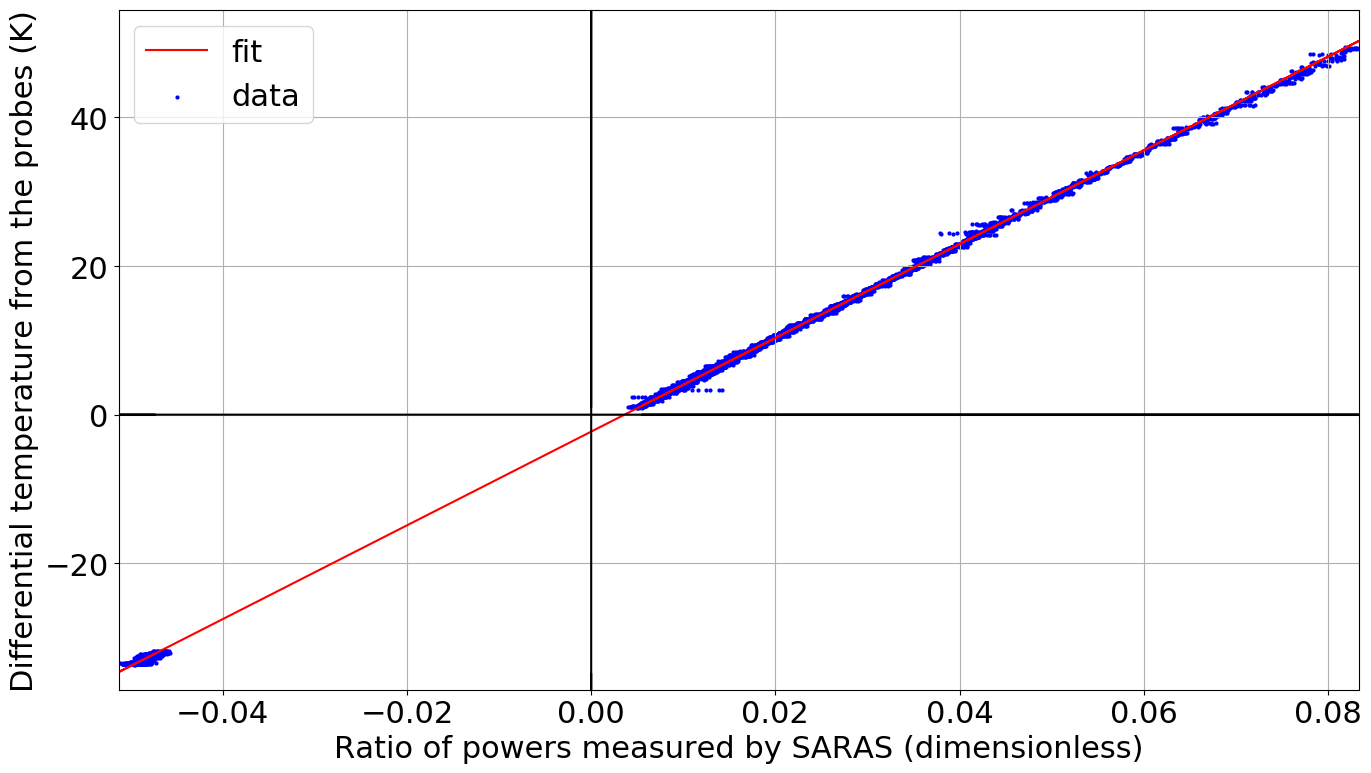}
\caption{Fit to data acquired with termination at antenna port placed in warm and cold baths, which provides estimate of the absolute calibration scale factor $T_{STEP}$.}
\label{fig:abs_cal}   
\end{center}   
\end{figure}

The equation to be solved is a linear equation requiring at least two measurements to solve for the unknowns.  Traditionally, the termination is placed in baths, the temperature of the bath set at two different temperatures and measurement data recorded, and the data used to exactly solve for  the two unknowns $T_{\rm STEP}$ and $T_{\rm os}$.  Instead, for the SARAS~3 receiver, we made two dynamic measurements.  The termination is first immersed in a hot water bath in a well insulated dewar and allowed to cool slowly over time. Separately, the termination in immersed in ice-cold water in the dewar and allowed to slowly warm over time.  The recorded data provide overdetermined solutions.  Plot of the differential of the probe temperatures:  $(T_{\rm 50,m} - T_{\rm REF,m})$ versus ratio of the differential powers recorded $(P_{\rm OBS} - P_{\rm REF})/(P_{\rm CAL} - P_{\rm REF})$ is shown in Fig.~\ref{fig:abs_cal}.  The fit of a straight line yields the slope and intercept, which are the two unknowns $T_{\rm STEP}$ and $T_{\rm os}$.

The fit gives the value of $T_{\rm STEP}$ to be 630~K and the y intercept gives $T_{\rm os}$ to be about $-2.3$~K.  The goodness of fit is a confirmation of the model for the experimental setup.

\subsection{Termination tests}
\label{subsec:term_test}

Eq.\ref{eq:Tinf_calib_main} gives a detailed description of the measured spectrum, including the expected systematic structures that it may contain owing to non-ideal component behavior within the receiver chain. The measured spectrum may be viewed on the whole as consisting of three components. The first is the signal from the antenna modified by the transfer function of the system.  Departure in this transfer function from an ideal flat response is a multiplicative error.  The measurement differences the antenna signal with the reference power.  The reference is the second component and we assume in the analysis herein that it is ideal and of flat spectrum to the accuracy required for CD/EoR detection. The third component is additive noise from along the receiver path: this is dominated by the receiver amplifier noise and the internal additives appear in the measured spectrum with their band shapes multiplied by corresponding transfer functions.   

In this section, we present results of laboratory measurements done as qualification tests of the SARAS~3 receiver, with the antenna replaced successively by precision terminations and a load that is a circuit simulator of the antenna characteristics. The load resembling the antenna characteristics---hereinafter called an `antenna simulator'---is a resistance-inductance-capacitance network purpose built to have a reflection coefficient $S11$ amplitude that matches the reflection coefficient measured for the antenna.  The aim of the qualification tests were to examine the limitations of the design effort in the SARAS~3 receiver, which was aimed at realizing a spectral radiometer whose unavoidable systematics were relatively smooth and hence separable from CD/EoR spectral profiles predicted in cosmological models.  

In order to investigate the internal systematics of the radiometer, ideally a measurement that does not contain any sky signal is desired. This can be obtained by replacing the antenna with a perfectly mismatched termination, which may be an electrical open or short.  Indeed, such terminations can provide an estimate of the maximum levels of additive systematics that any spectral measurement with the radiometer might contain.  Data for such a test were acquired by replacing the antenna with precision open and short terminations and acquiring spectra with the full SARAS~3 receiver chain for about 16~hr for each termination, with the receiver cycling through the switch states exactly as designed for celestial observations. The data were then calibrated in the standard process and reduced to provide a single average spectrum for each termination.  The terms in the measurement equation that depend on the reflection coefficient $\rm \Gamma_A$ at the antenna terminal will flip sign when changing from an electrical short to open at the antenna terminals.  Therefore, we examine the measurement data linearly combined to yield (open-short)/2 and (open+short)/2, which would separate terms that depend on $\rm \Gamma_A$ and those that do not, respectively.   

\subsubsection{Modeling laboratory measurements with the measurement equation}
\label{subsec:meas_equ_fit_data}

The measurement equation, given by Eq.~\ref{eq:Tinf_calib_main}, was fitted to the measured data in two frequency bands 50--100~MHz and 90--180~MHz, which roughly correspond respectively to the bands in which CD and EoR related 21-cm signals are expected.  The reflection coefficients of the precision open and short terminations are assumed to be ideal, since at the frequencies of interest the effects of their fringing capacitance and inductance are negligible. The fraction $f$ of the receiver noise voltage that emerges from the input of the amplifier and back propagates towards the antenna is modeled as a complex variable that is a constant, independent of frequency, throughout the band of interest.  Similarly, the complex reflection coefficient at the input terminals of the receiver, $\rm \Gamma_N$, is also modeled as a complex variable that is a constant throughout the band of interest. The path length $l$ is a free parameter in the modeling.   Spectra recorded with the switch connecting the receiver to the REF port was calibrated using the difference CAL$-$REF; this measurement is expected to represent the sum of receiver noise and ambient temperature of the matched load at the REF port.  We subtract the recorded ambient temperature of the REF termination from this calibrated REF spectrum and derive an estimate of the receiver noise temperature $\rm T_N$ by fitting a third-order polynomial to this residual calibrated REF port spectrum.  Thus we effectively adopt a five-parameter model for the system---two complex variables and one real variable---and fit this to the termination measurement data.   Suitable boundary conditions are imposed to prevent the fits from returning unphysical parameters.

\begin{figure}[htbp]
\centering
\begin{minipage}[!ht]{1.0\linewidth}
\includegraphics[width=\linewidth]{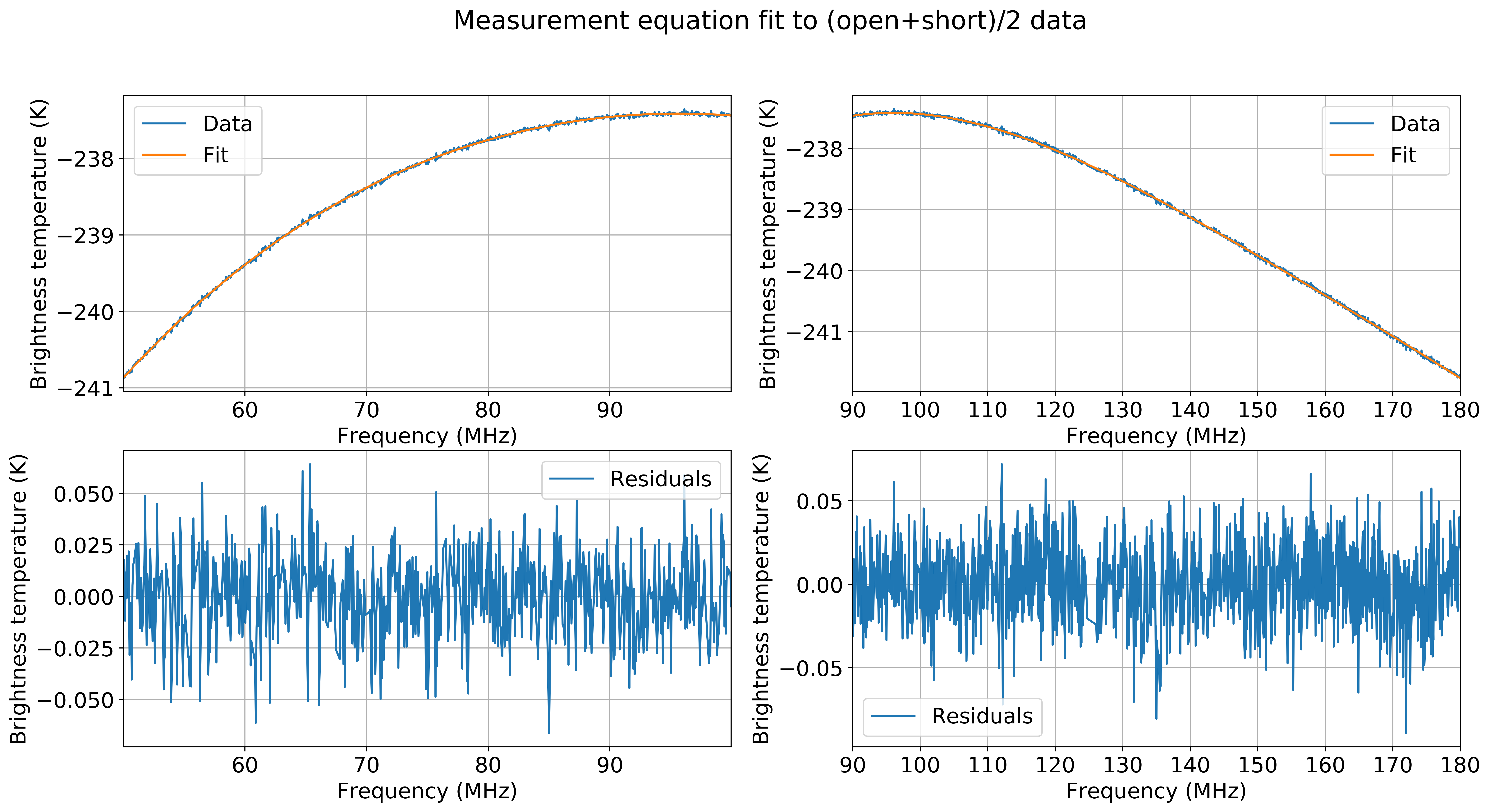}
\vspace{3mm}
\end{minipage}
\begin{minipage}[!ht]{1.0\linewidth}
\includegraphics[width=\linewidth]{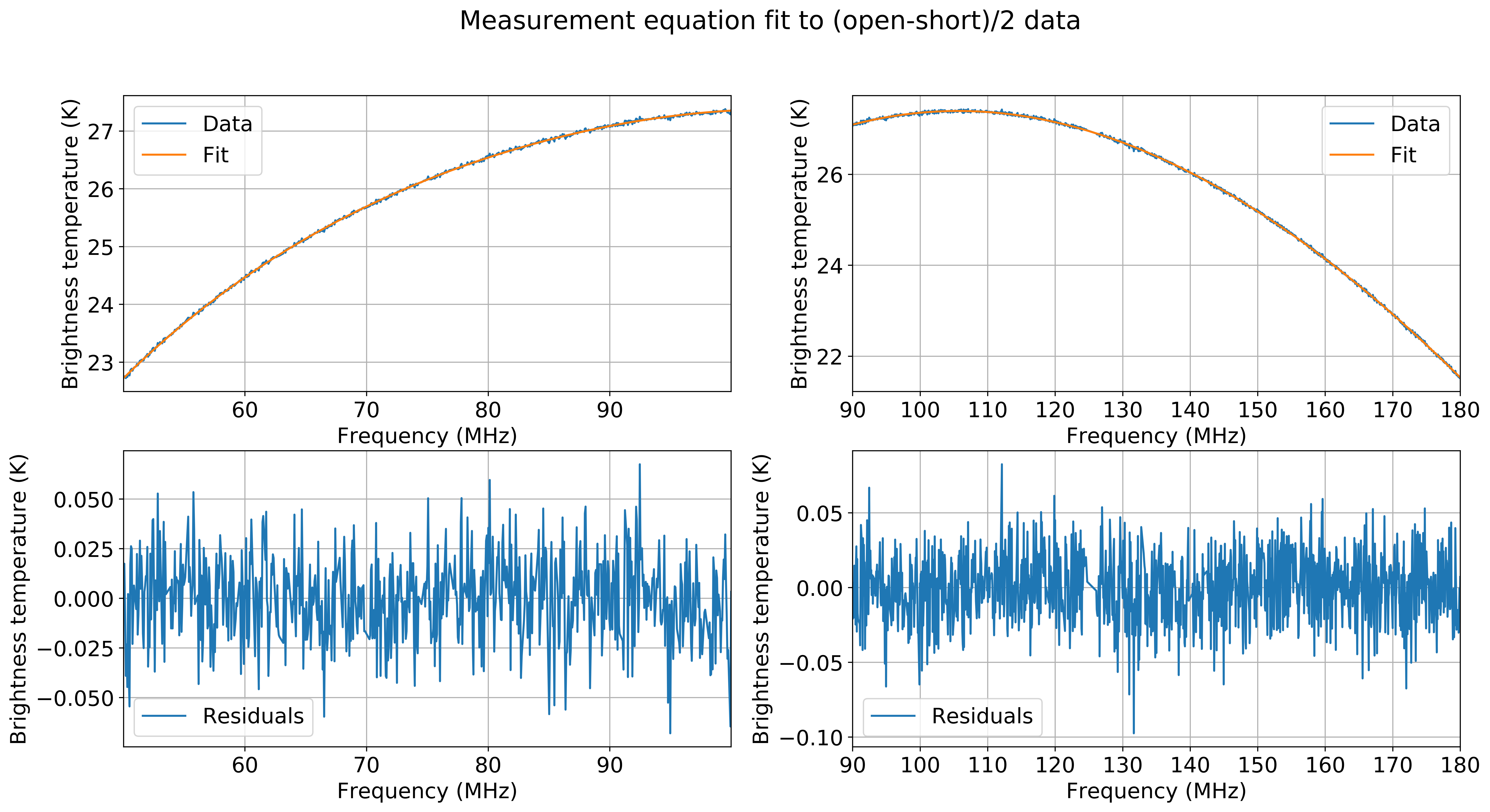}
\end{minipage}
\caption{Modeling measured data in the 50--100~MHz and 90--180~MHz bands using the measurement equation Eq.~\ref{eq:Tinf_calib_main}.  The sum and difference of data acquired with the antenna replaced with precision open and short terminations are modeled. }       
\label{fig:meas_equ_fit}
\end{figure}

In Fig.~\ref{fig:meas_equ_fit}, the results of fitting Eq.~\ref{eq:Tinf_calib_main} to (open$+$short)/2 and (open$-$short)/2 are shown. It is seen that the measurement equation is indeed successful in modeling the measured spectra and its various components to within the measurement noise. However, in order to accurately model the system response to mK levels the assumptions made above regarding various model parameters have to be relaxed. The various model parameters have their own frequency dependence and if we were to attempt an accurate estimation of that as well from the measured data, the optimization problem would have a large number of free parameters.   Allowing for larger numbers of free parameters in the measurement equation would result in a model that would fit the data with reduced residuals. However, the model describing the systematics would also fit out any CD/EoR signal when used to model sky spectra, and hence substantially reduce the sensitivity of the radiometer.  Alternately,  $\rm \Gamma_N$ and $f$ may be measured in the laboratory or in-situ using, for example, an accurate impedance tuner at the receiver input to obtain data for various impedance states (this is sometimes referred to by the name source pulling).  However, such measurements are difficult to make with the desired accuracy because the receiver would have to be switched to a different measurement apparatus for this, and the receiver parameters might change over time and between laboratory conditions and the field. Therefore, we model the laboratory---and field---data with maximally smooth polynomials \cite{2017ApJ...840...33S}, as discussed below.

\subsubsection{Modeling laboratory measurements with maximally smooth polynomials}
\label{subsec:max_smooth_fit_data}

The measurement data with different terminations, and in bands appropriate for detection of CD and EoR features, were fitted using maximally smooth polynomials.  In its smoothest formulation, maximally smooth polynomials allow a single maximum or minimum within the band.  A relaxation is to allow for a single zero crossing within the band in the second and higher order derivatives, resulting in a modified maximally smooth polynomial.  These polynomials may be of arbitrarily large order; nevertheless, they would not completely fit out CD/EoR signals that might be present in measurements of sky spectra.  The maximally smooth polynomials would fit out part of the signals being searched for, and hence reduce the sensitivity of the experiment. Modified maximally smooth polynomials would fit out a greater fraction.  The goal of the receiver qualifying tests has been to evaluate whether the internal systematics may be modeled with maximally smooth functions, or in its modified form. 

\begin{figure}[htbp]
\centering
\begin{minipage}[!ht]{0.8\linewidth}
\includegraphics[width=\linewidth]{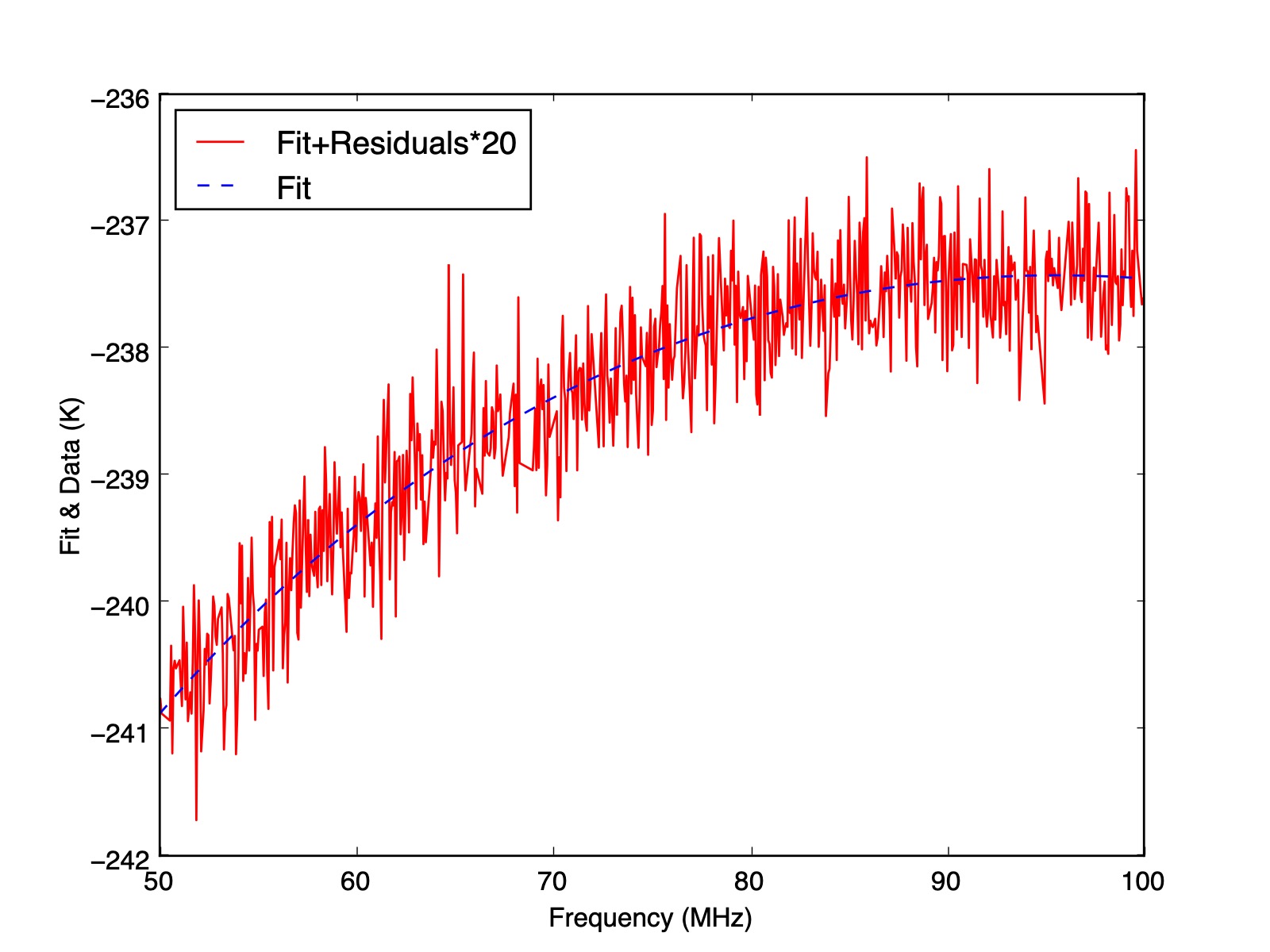}
\end{minipage}
\begin{minipage}[!ht]{0.8\linewidth}
\includegraphics[width=\linewidth]{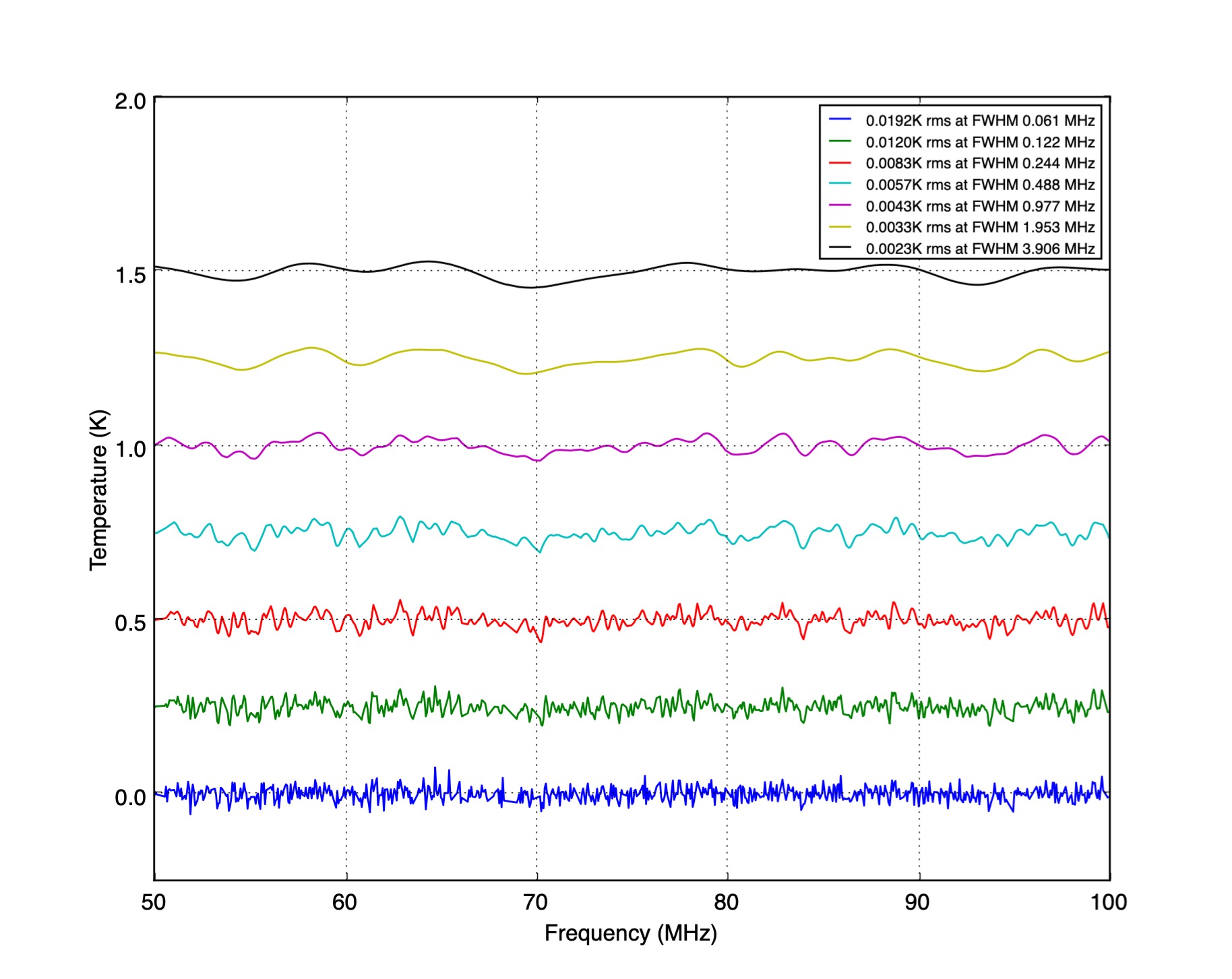}
\end{minipage}
\begin{minipage}[!ht]{0.8\linewidth}
\includegraphics[width=\linewidth]{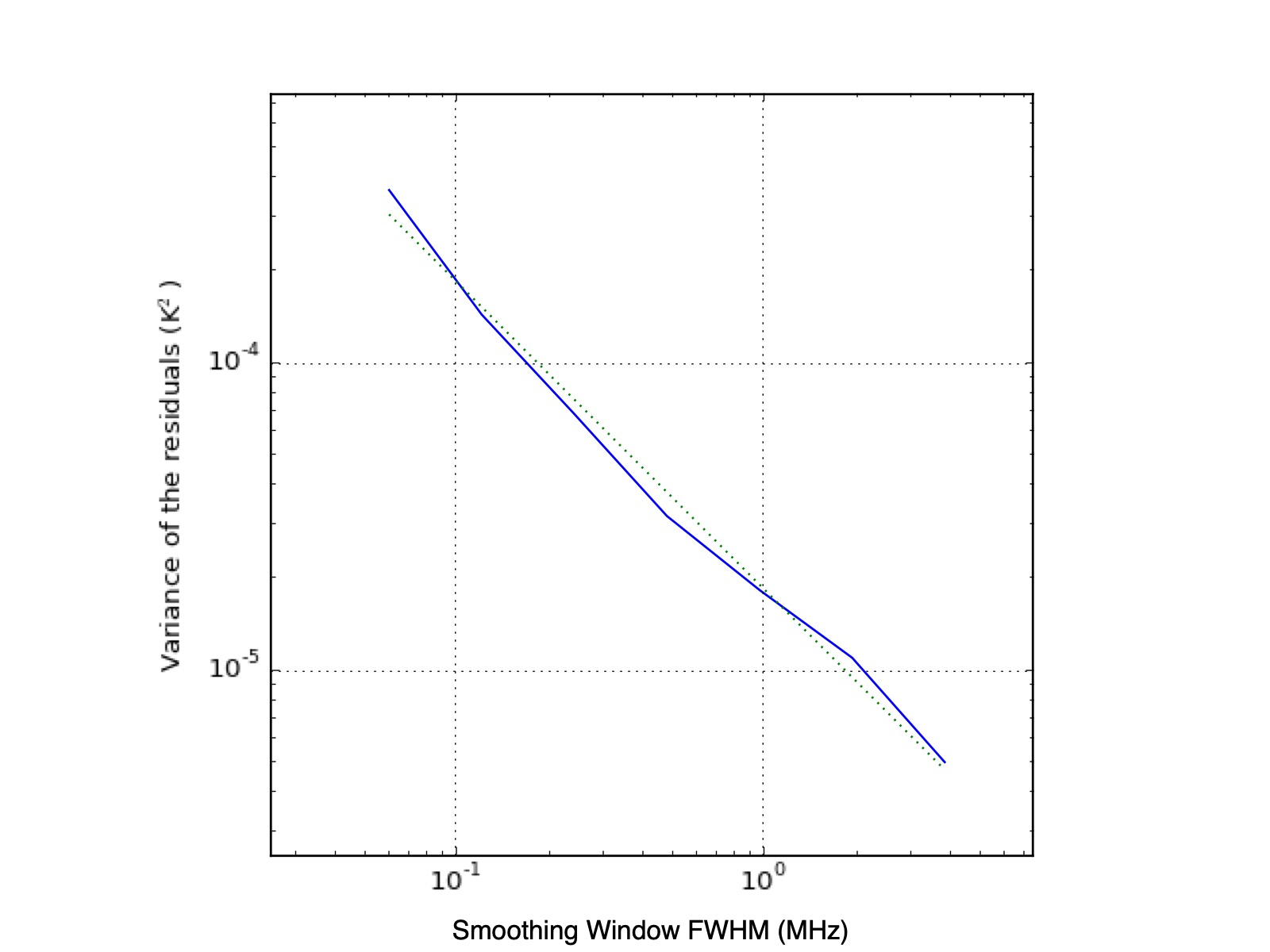}
\end{minipage}
\caption{The result of modeling SARAS~3 systematics in the CD band 50--100~MHz.  The sum of measurement data acquired with precision electrical open and short terminations at the antenna terminals is fitted using a maximally smooth polynomial form.  The top panel shows the measurement data and fit together; the residuals are magnified by factor 20 for clarity. The middle panel shows the fitting residuals smoothed using kernels of increasing fwhm.  The data with native resolution of 61~kHz is the lowest trace and spectra smoothed progressively to larger fwhm are shown above that with offsets of 0.25~K; traces are magnified by factors that keep the apparent rms the same on all smoothing.  The legend in this middle panel lists the rms at different spectral resolutions. The continuous line in the bottom panel shows the run of variance in the residuals versus spectral resolution; the expected rate of fall in noise with smoothing is indicated by the dotted line.}       
\label{fig:ops_fit_50100}
\end{figure}

In Fig.~\ref{fig:ops_fit_50100}, we fit the (open$+$short)/2 measurement data versus frequency with a {\it maximally smooth} polynomial  in the CD band 50--100~MHz.   This linear combination would cancel terms in Eq.~\ref{eq:Tinf_calib_main} that depend on odd powers of $\Gamma_A$; therefore, the last term in the third line of the equation survives whereas part of the series in the second line drops out.  The panel on the top shows the data with MS fit overlaid, demonstrating the goodness of fit.  The measurement data is with negative temperatures in the y-axis because the measurement represents difference between powers from the antenna terminal and that from the reference. For this measurement, an electrical short/open is at the antenna terminals and this has lower noise power compared to that from the reference termination, which is a matched ambient temperature load. The middle panel shows the fitting residuals.  The residuals are displayed with their native resolution of 61~kHz and also shown smoothed using kernels of increasing bandwidths. Using Eq.~\ref{eq:del_Tmeas_eq} and adopting realistic values for the noise temperatures in the different switching states, as discussed in Sec.~\ref{sec:sensit}, we expect that the calibrated (open$+$short)/2 spectrum would have an rms noise of 20~mK at native resolution, which is indeed what is measured for the data. The bottom panel of the figure shows the variation in the variance of the residuals as a function of the full width at half maximum (fwhm) of the smoothing kernel.   If the spectra were Gaussian random noise, this variation is expected to be a straight line with slope $-1$ in log-log domain; for comparison, we also show in the panel this expected rate of fall.  

\begin{figure}[htbp]
\centering
\begin{minipage}[!ht]{0.9\linewidth}
\includegraphics[width=\linewidth]{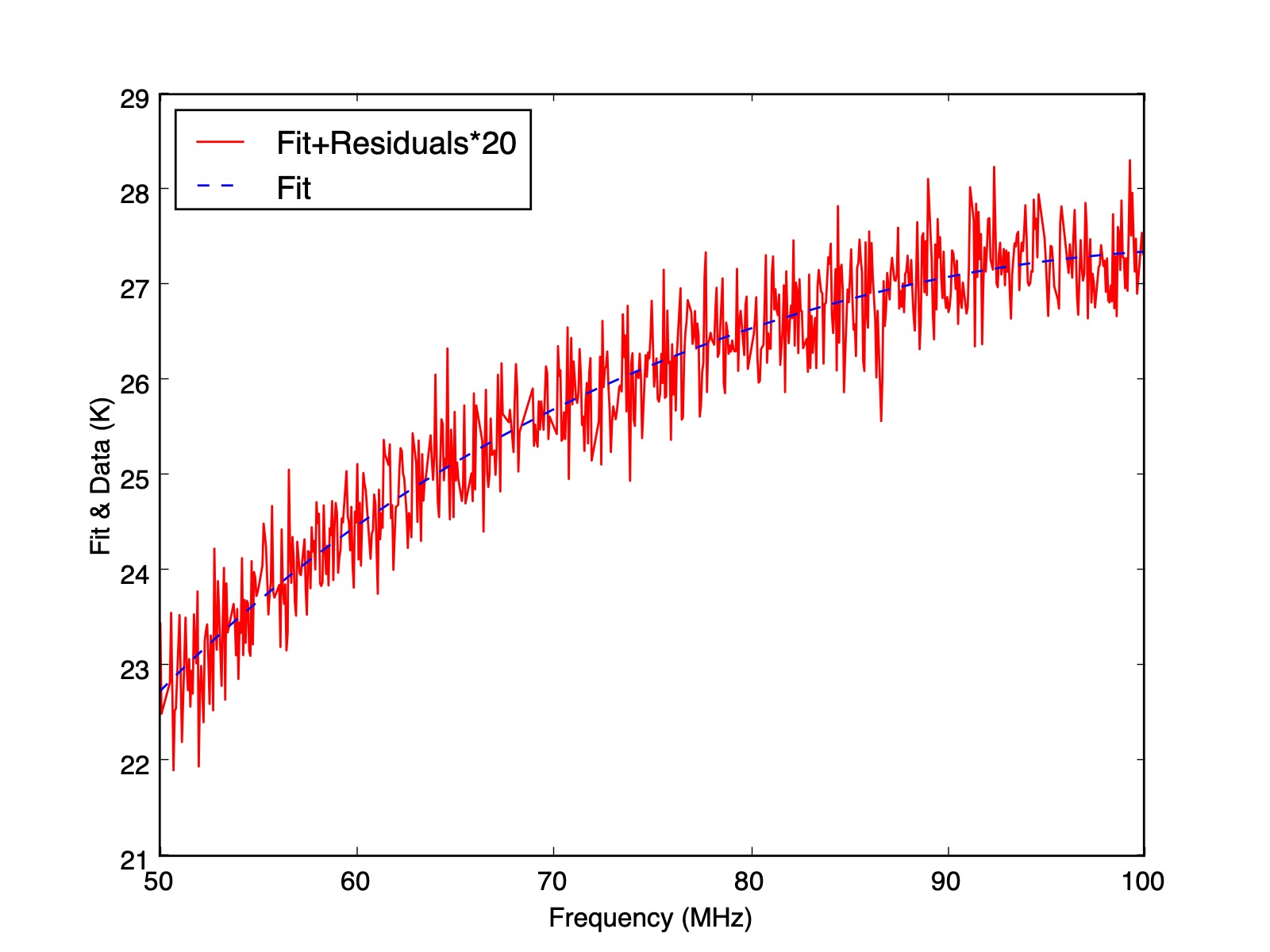}
\end{minipage}
\begin{minipage}[!ht]{0.9\linewidth}
\includegraphics[width=\linewidth]{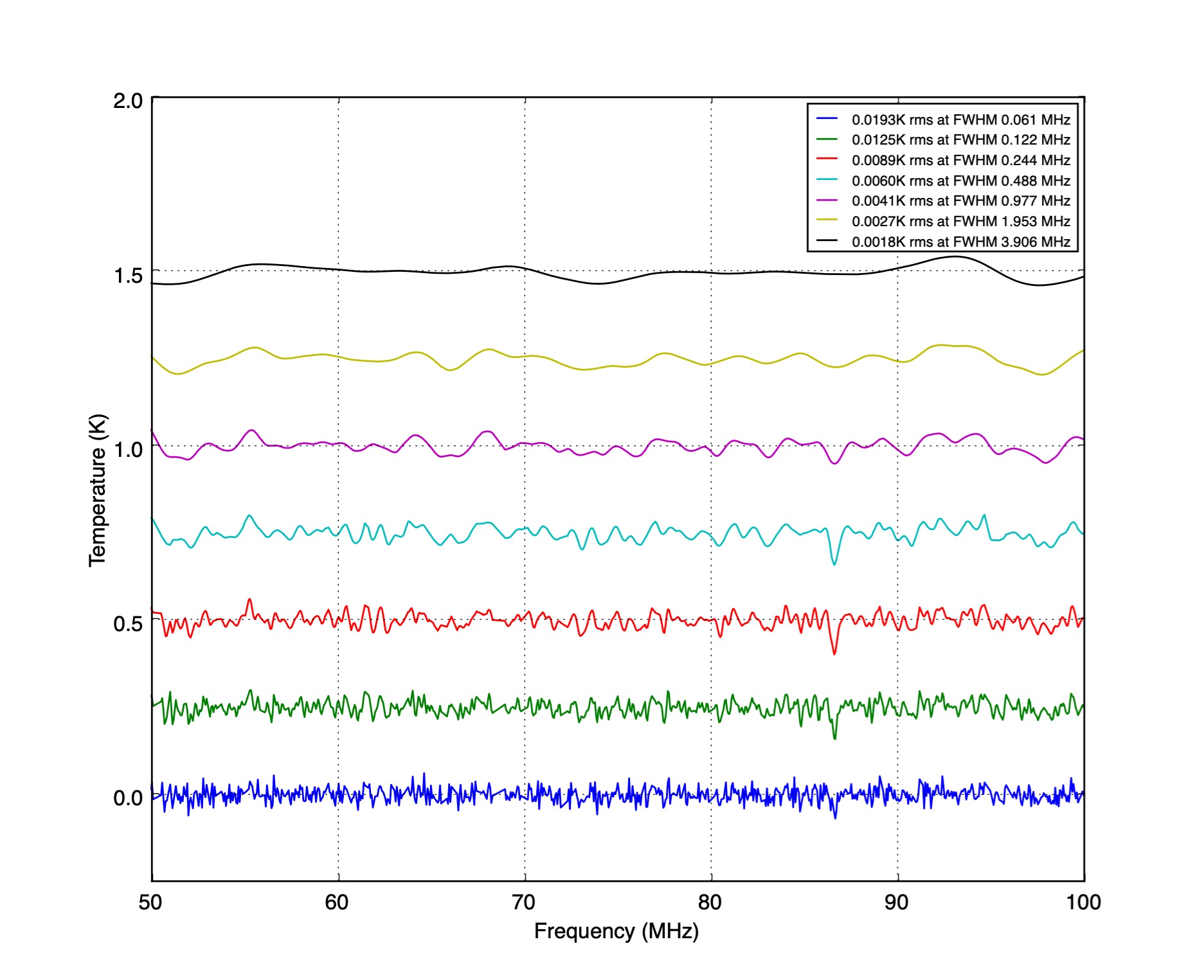}
\end{minipage}
\begin{minipage}[!ht]{0.8\linewidth}
\includegraphics[width=\linewidth]{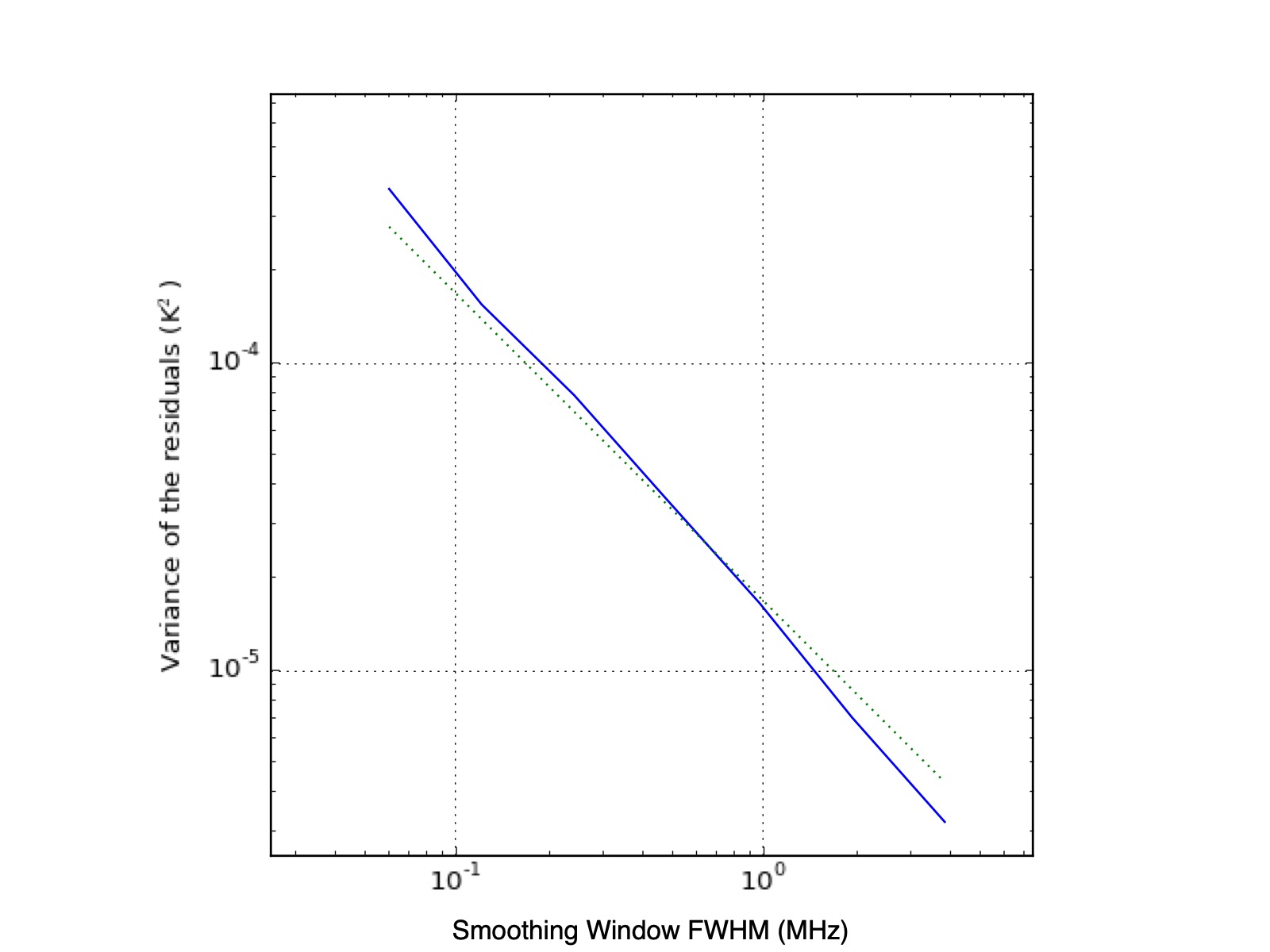}
\end{minipage}
\caption{Modelling measurement data in the 50--100~MHz CD band.  Here the difference of data acquired with open and short terminations at the antenna terminals is modeled.  The three panels depict analyses same as that in the previous figure.}       
\label{fig:oms_fit_50100}
\end{figure}

Fig.~\ref{fig:oms_fit_50100} shows the result of fitting an MS polynomial to the (open$-$short)/2 spectrum in the 50--100~MHz band. This linear combination is expected to cancel the entire term in the third line of Eq.~\ref{eq:Tinf_calib_main}, which wholly depends on even powers of $\Gamma_A$, and also part of the series in the second line where the terms depend on even powers of $\Gamma_A$.   As in the case of the analysis of the (open$+$short)/2 spectra, here too we have displayed the fits, fit residuals smoothed to lower resolutions, and compared the run of variance with smoothing fwhm.  

\begin{figure}[htbp]
\centering
\begin{minipage}[!ht]{0.9\linewidth}
\includegraphics[width=\linewidth]{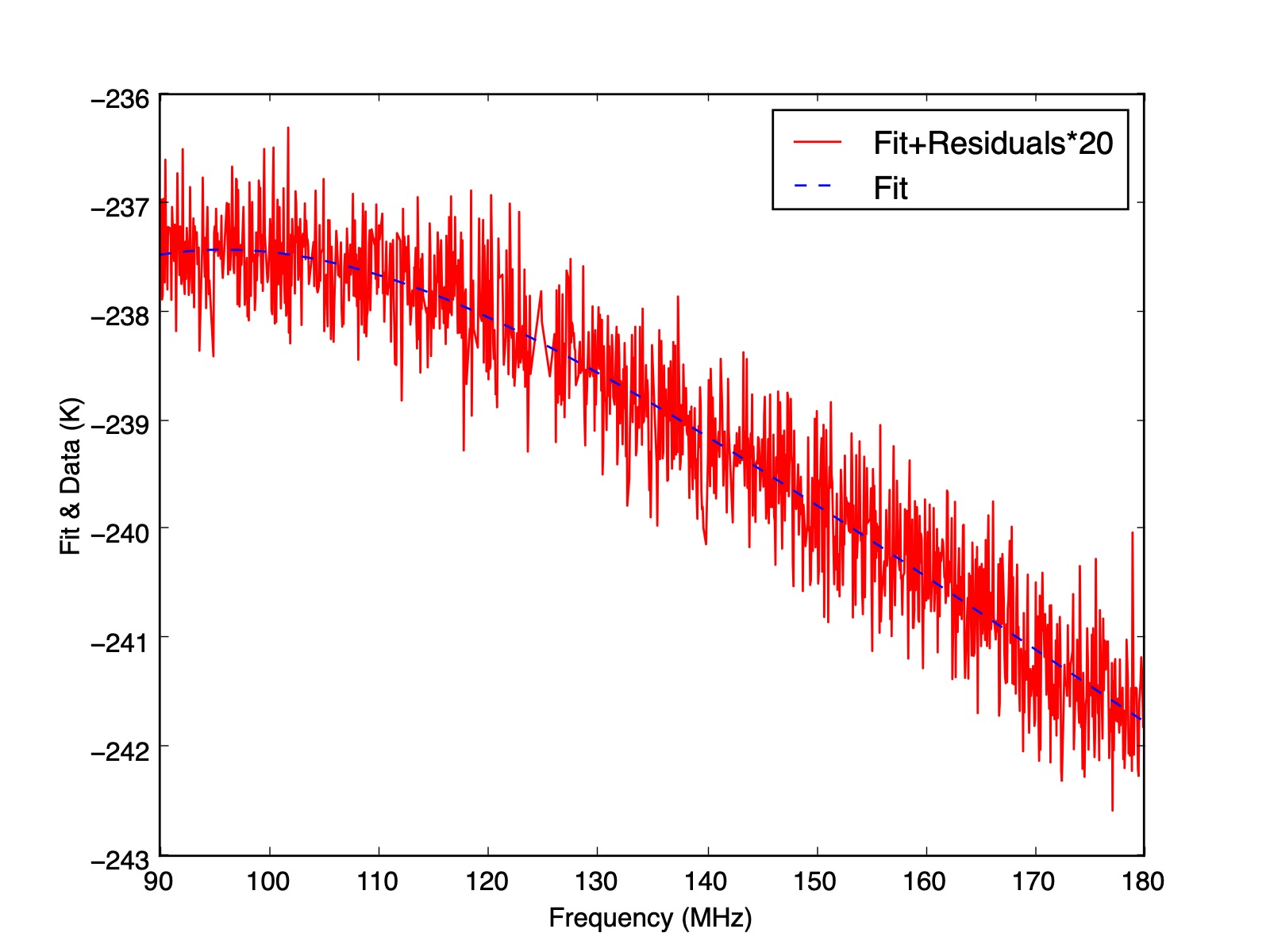}
\end{minipage}
\begin{minipage}[!ht]{0.9\linewidth}
\includegraphics[width=\linewidth]{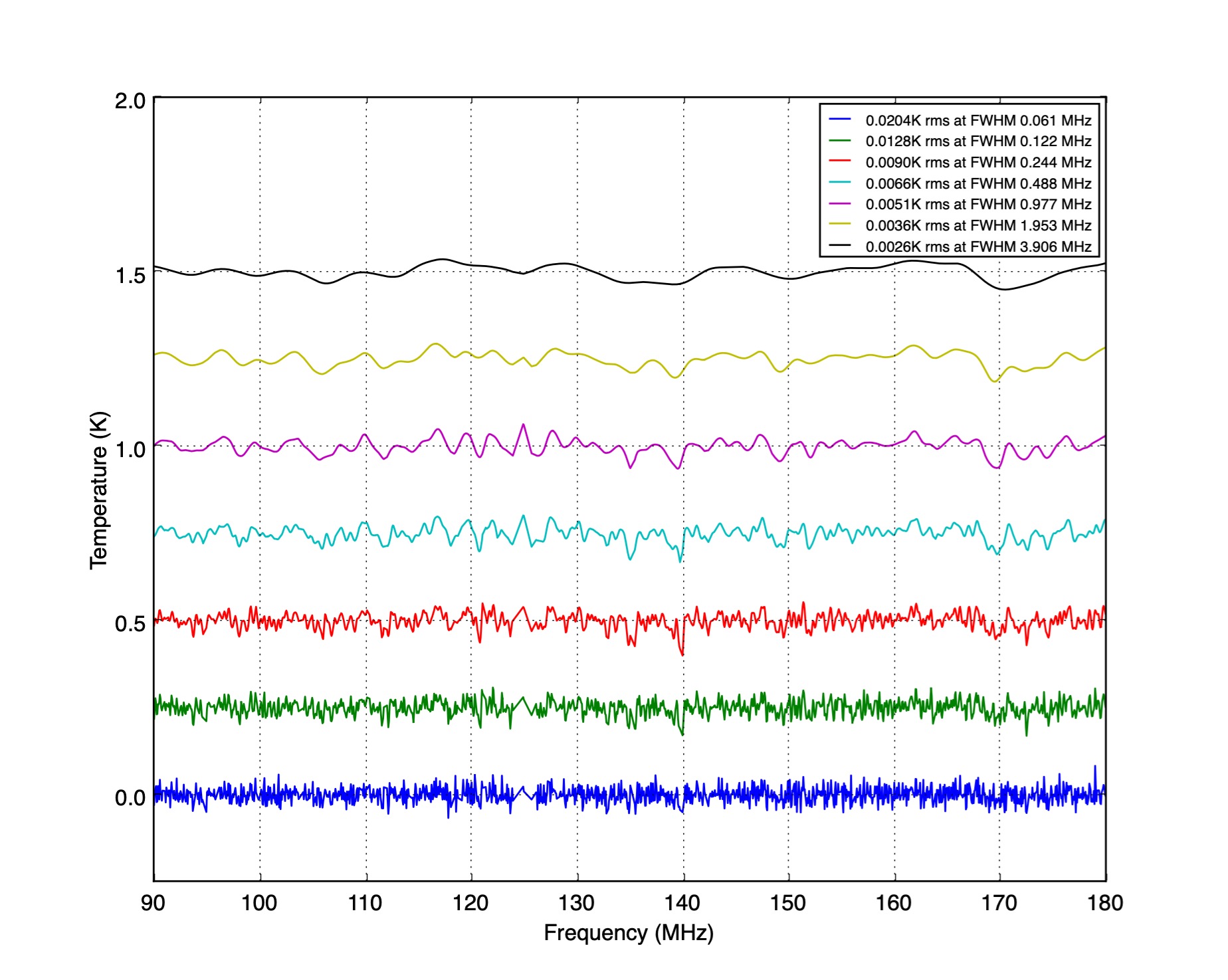}
\end{minipage}
\begin{minipage}[!ht]{0.8\linewidth}
\includegraphics[width=\linewidth]{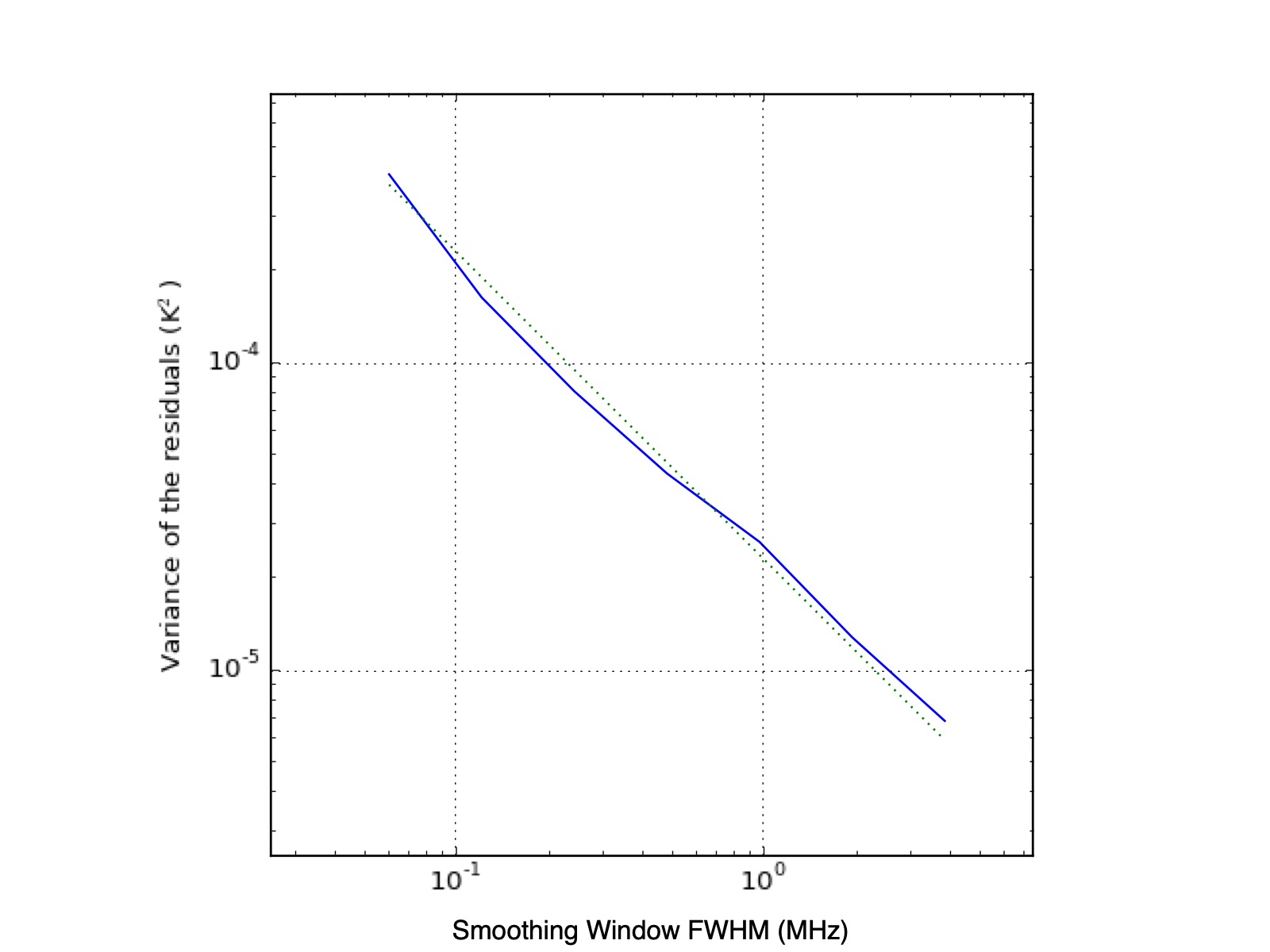}
\end{minipage}
\caption{Modeling laboratory measurement data in the EoR band 90--180~MHz.  In this figure, the sum of calibrated spectra acquired with open and short terminations at the antenna terminals is modeled.   The three panels depict analyses same as that in the previous figure; however, for the wider EoR band the data was modeled using the modified form of the maximally smooth function described in the text, which allows for one zero crossing in higher order derivatives. }       
\label{fig:ops_fit_80160}
\end{figure}

\begin{figure}[htbp]
\centering
\begin{minipage}[!ht]{0.9\linewidth}
\includegraphics[width=\linewidth]{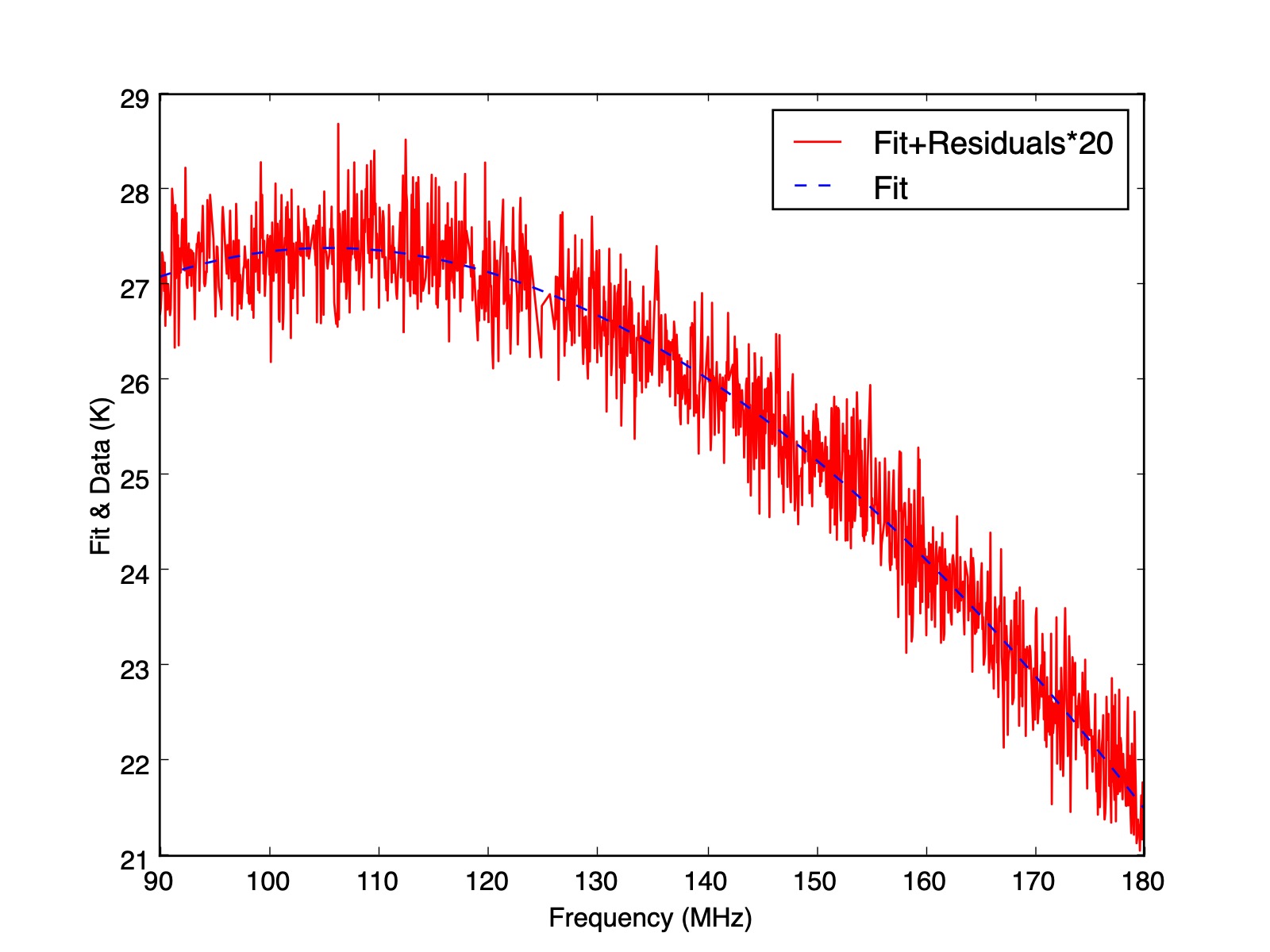}
\end{minipage}
\begin{minipage}[!ht]{0.9\linewidth}
\includegraphics[width=\linewidth]{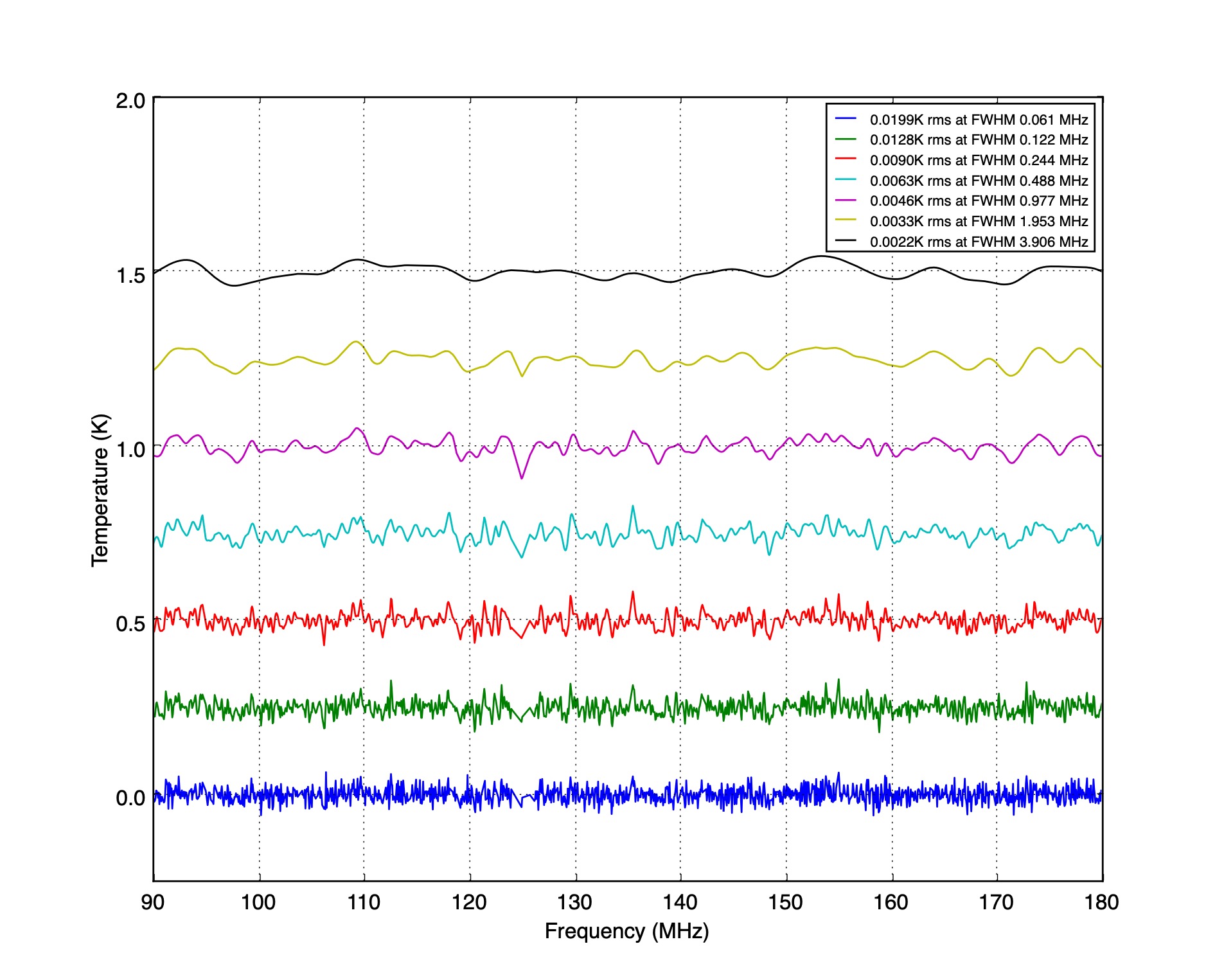}
\end{minipage}
\begin{minipage}[!ht]{0.8\linewidth}
\includegraphics[width=\linewidth]{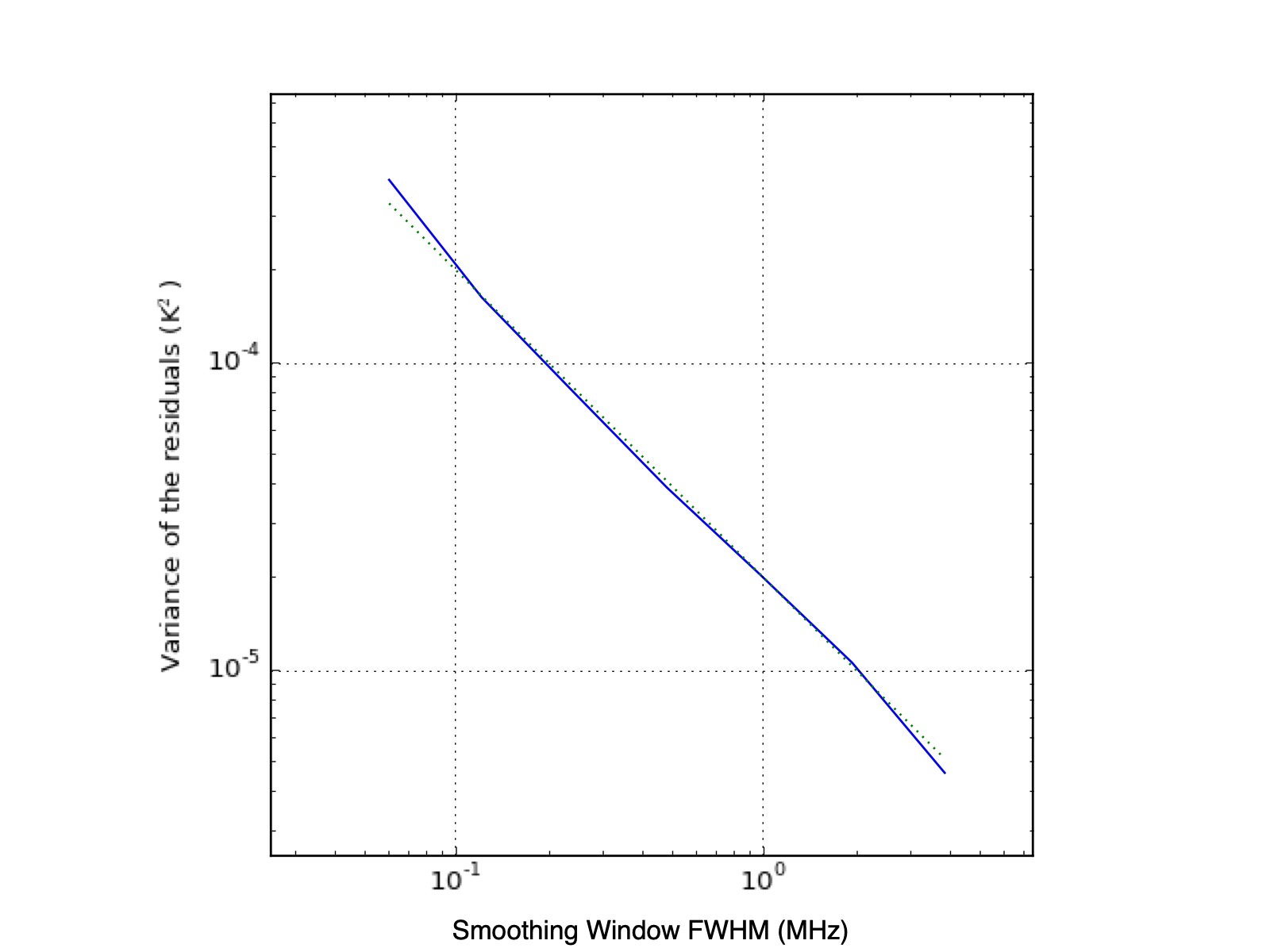}
\end{minipage}
\caption{Modeling for systematics in the EoR band 90--180~MHz.  In this figure, data that is the difference of those acquired with open and short terminations at the antenna terminals is examined. Here also, as in the previous figure, the fitting function is a modified maximally smooth function. The three panels depict analyses same as that in the previous figure.}       
\label{fig:oms_fit_80160}
\end{figure}

Figs.~\ref{fig:ops_fit_80160} and \ref{fig:oms_fit_80160} show the results of modeling (open$+$short)/2 and (open$-$short)/2 measurement data respectively in the EoR band 90--180~MHz.   The increased bandwidth in these cases necessitated the use of the modified MS polynomial discussed above.

\begin{figure}[htbp]
\centering
\begin{minipage}[!ht]{0.9\linewidth}
\includegraphics[width=\linewidth]{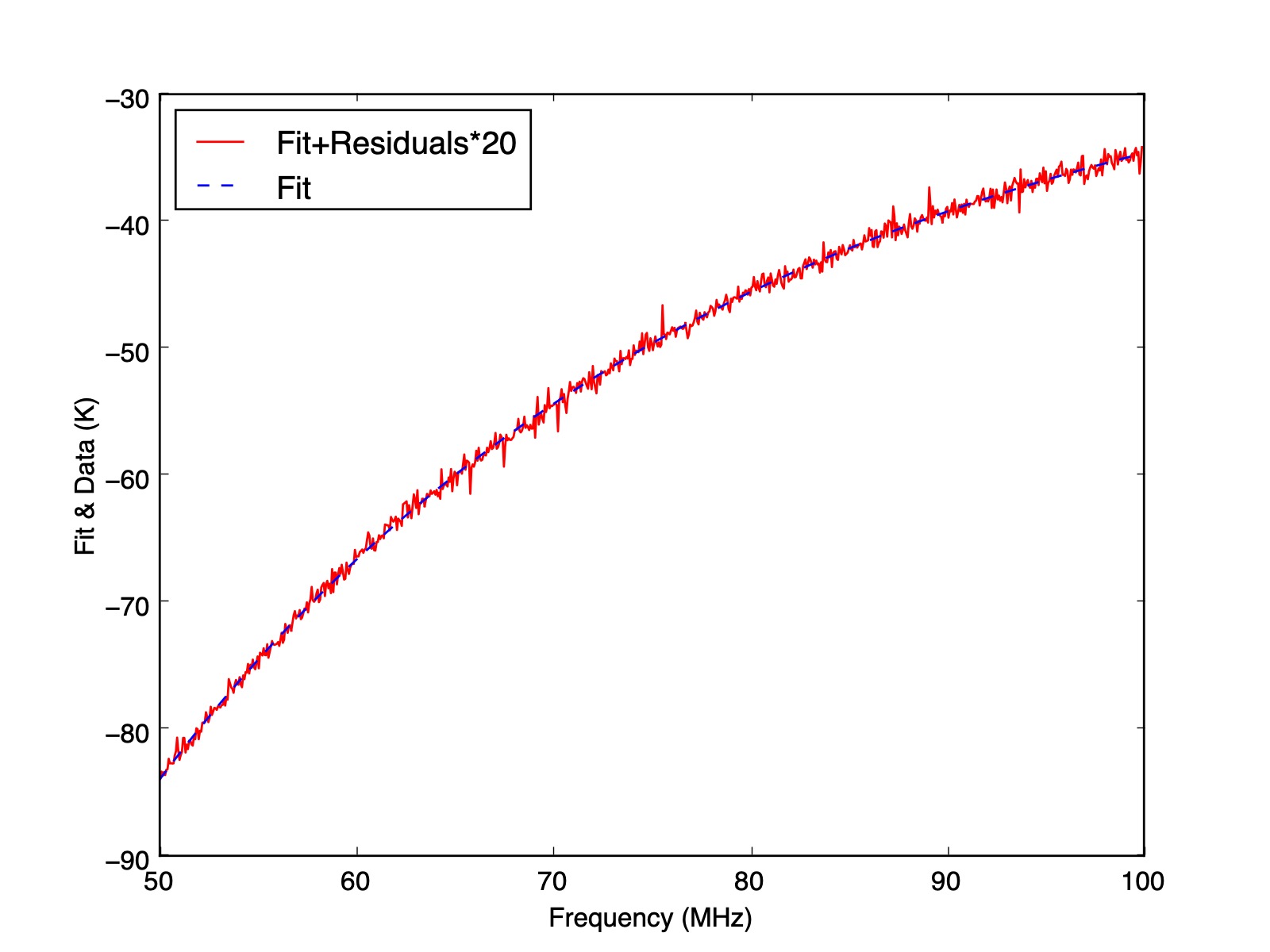}
\end{minipage}
\begin{minipage}[!ht]{0.9\linewidth}
\includegraphics[width=\linewidth]{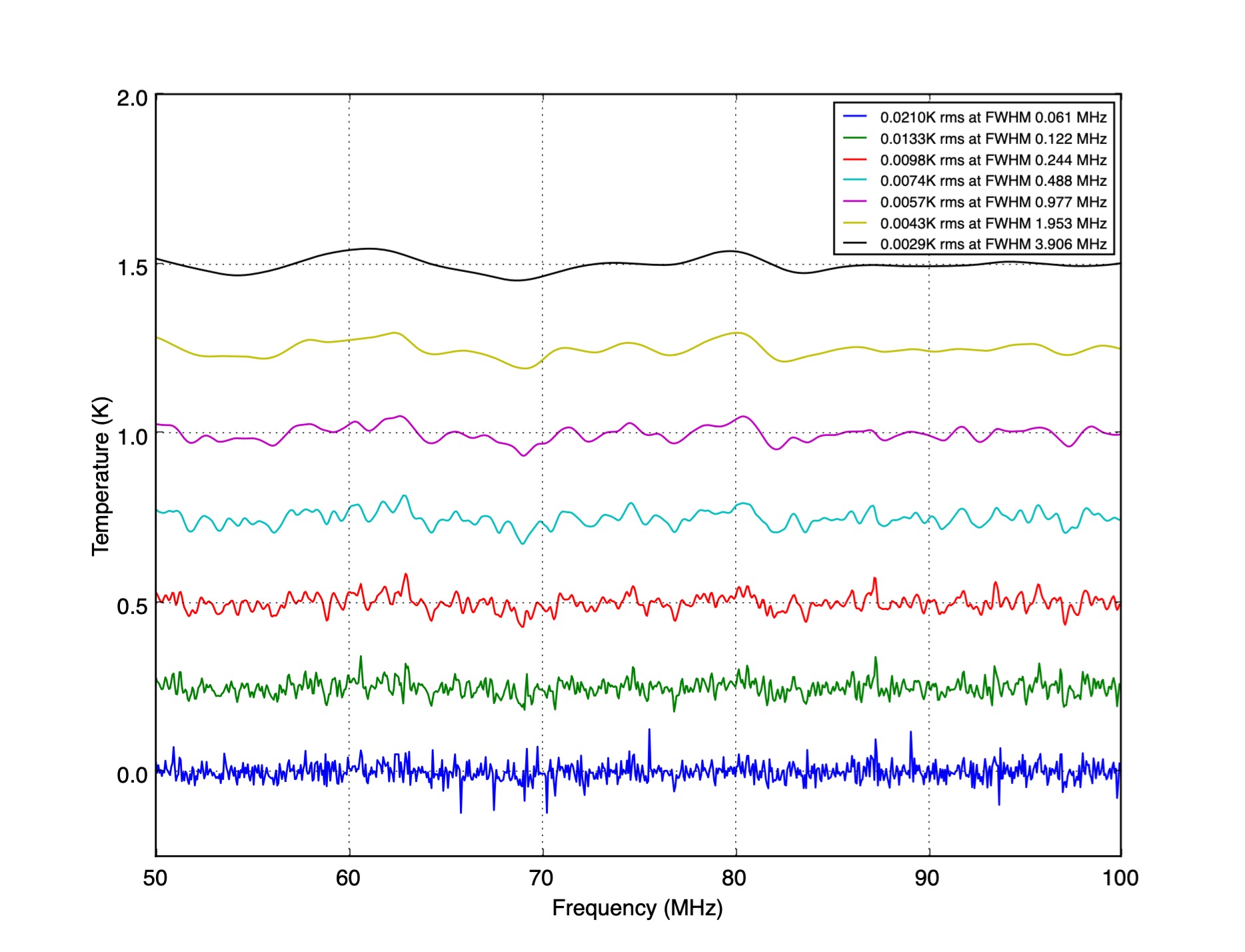}
\end{minipage}
\begin{minipage}[!ht]{0.8\linewidth}
\includegraphics[width=\linewidth]{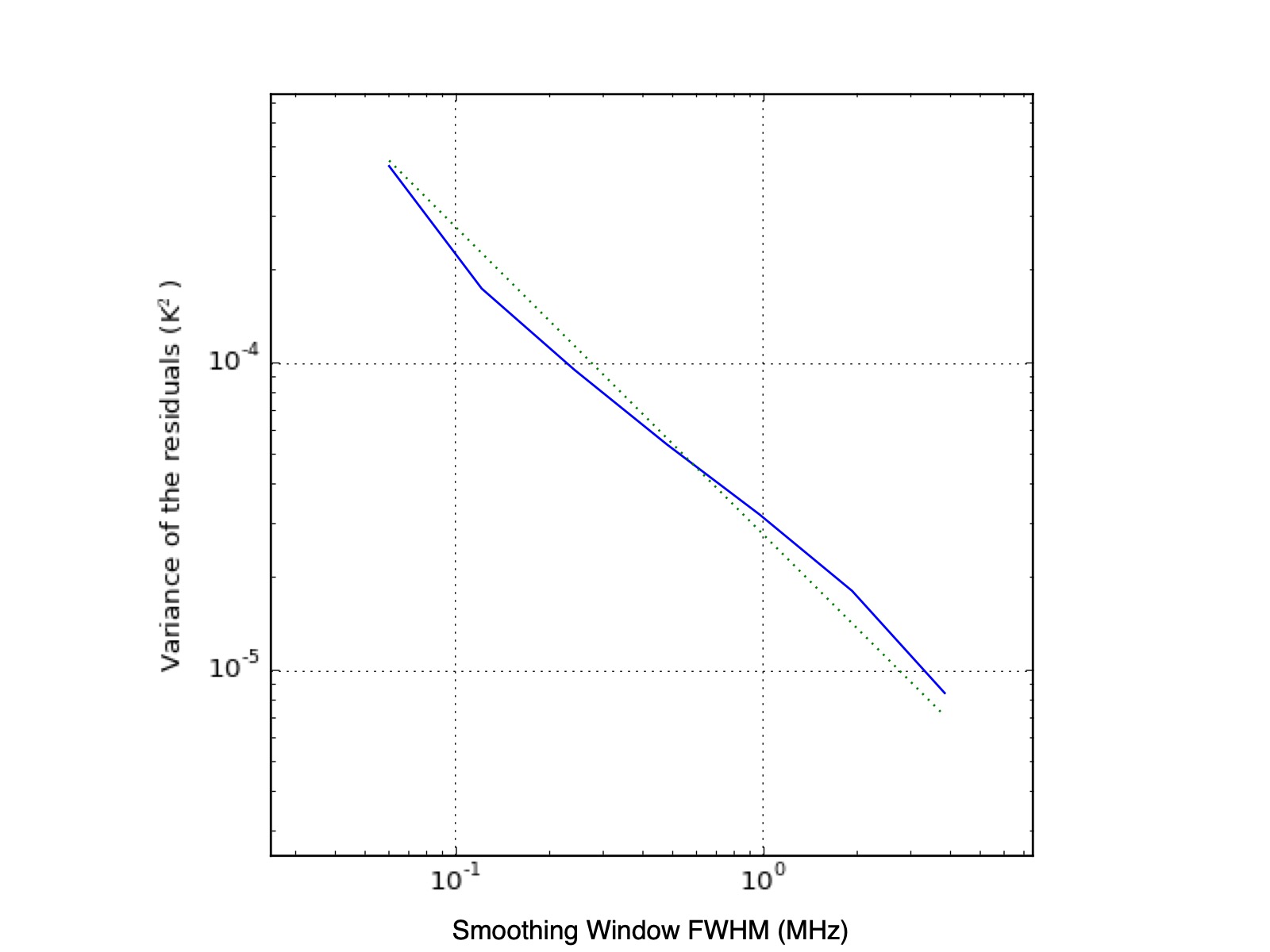}
\end{minipage}
\caption{Modeling measurement data acquired with a 1-port antenna circuit simulator replacing the antenna. The three panels depict analyses same as that done in the previous figure.}       
\label{fig:RLC_fit}
\end{figure}

In order to evaluate the internal systematics in a setup in which their characteristics would be similar to that while observing with the SARAS~3 antenna \cite{2020ITAP....Raghu}, we examine the measurement data in a third case where the antenna is replaced with a circuit simulator: the antenna simulator discussed above.  The resistive component of the antenna simulator is at ambient temperature, hence the effective antenna temperature in this case would be the product of the ambient temperature and a ``reflection efficiency" for the simulator that is related to the reflection coefficient $S11$ of the 1-port network.   Additionally, the measurement data would be expected to reveal any systematics that result from system temperature components suffering internal reflections at the antenna terminals.  Results that follow from a maximally smooth fit to the measurement data acquired in this configuration are shown in Fig.~\ref{fig:RLC_fit}.  Once again, due to the subtraction of the reference temperature from the antenna temperature in the computation of the difference measurement, the values in the y-axis are negative. 

In all the cases considered above, the fits of maximally smooth functions to the measurement data are good.  There are no obvious systematics in the residuals, whose rms decreases with smoothing to within a few mK.  This rms decreases with smoothing without any indication that systematics would limit the sensitivity to above mK.  Indeed the maximum deviation of the rms from that expected for Gaussian noise suggests that any residual systematic, that cannot be modeled with MS functions, has rms less than 1.2~mK.  The laboratory test measurements demonstrate the receiver fidelity and qualifies the receiver for experiments aiming to detect CD/EoR structures that are distinct from maximally smooth functions and with signal amplitudes above a mK.  

\section{Summary}
\label{sec:summary}

In this paper, we have presented the design philosophy, design, and performance of the SARAS~3 radiometer receiver and evaluated the capability of the system to detect the global 21~cm cosmological signal from cosmic dawn and reionization. We have discussed the new features in this receiver system that represent an improvement from earlier versions of the experiment. The system has been designed in such a manner so that inevitable systematic features---additive and multiplicative---that survive standard calibration would be maximally smooth and hence separable from CD/EoR signals. Double differencing is employed, switching the receiver between the antenna and a reference and phase switching to cancel additive spurious entering the signal path. Optical isolation is provided and a digital correlation spectrometer is used. The receiver is built to be compact and located at the terminals of the antenna.

The signal path in the receiver has been analyzed in detail, leading to derivation of a measurement equation that includes spectral contributions from multi-order reflections occurring between the receiver and the antenna, within the signal path. We have also presented estimates of the sensitivity of the system taking into account the different system temperature components and the measurement equation, arising from the double differencing and the proposed switching scheme for calibration.   

Finally, we report the qualification tests carried out in the laboratory to experimentally confirm that any systematics present in the data acquired with this system will not hinder a detection of 21~cm signals from CD/EoR. Using data acquired, with the antenna successively replaced with precision electrical terminations and an antenna simulator network, we have demonstrated that the system has no residual additive systematics above 1~mK.  Thus cosmological signals received by the SARAS~3 antenna and processed by the receiver would appear in measurement data without suffering confusion above 1~mK.  

The SARAS~3 receiver design and implementation is adequate for the detection of 21-cm signals predicted in standard models for cosmic dawn and reionization.

\begin{acknowledgements}
We thank RRI Electronics Engineering Group, particularly Kasturi S., for their assistance in receiver assembly. We also thank the Mechanical Engineering Group (RRI), led by Mohamed Ibrahim, for construction of chassis and shielding cages for receivers. Gaddam Sindhu took an active role in implementing the receiver. We are grateful to the staff at the Gauribidanur Field Station led by H.A. Ashwathappa for providing excellent support in carrying out field tests and measurements.
\end{acknowledgements}

\begin{appendices}
\section{Derivation of the measurement equation for SARAS~3}
\label{app:measu_equ}

The Dicke switch alternately connects the radiometer receiver to the antenna and to a reference load.  The reference load is a noise source followed by an attenuation, so that the reference noise temperature may be switched between ambient and high temperature states depending on whether the noise source is on or off, while maintaining the impedance of the reference constant.

\begin{figure}[h!]
  \begin{center}
    \begin{circuitikz}
      \draw (0, 0)
      node[antenna]{};
      \draw (0,0)
      to[short, f=$l$] (2,0)
      to[amp] (4,0);
      \draw (0.5,1)
      node[label={above:$\rm V_A$}] {};
      \draw (-0.4,1)
      node[label={above:$\rm \Gamma_{A}$}] {};
      \draw (2,0)
      node[label={above:$\rm \Gamma_{N}$}] {};
      \draw (2.9,0.4)
      node[label={above:$\rm G, V_N$}] {};
      \draw (2,0)
      to[short, f=$\rm f$] (0,0);
      \end{circuitikz}
    \caption{Simplified noise model for the SARAS radiometer, when connected to the antenna.}
    \label{fig:rad_mod}
  \end{center}
\end{figure}
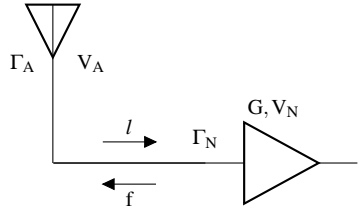

We first consider the case in which the Dicke switch is connected to the antenna and define the following terms to describe the noise model:
\begin{itemize}
  \item $\rm Z_{N}$ is the input impedance of the low noise amplifier (LNA) and $\rm \Gamma_{N}$ is the reflection coefficient of the LNA as referred to a $\rm Z_{0}$ = 50 $\rm \Omega$ measuring system impedance.
  \item $\rm Z_{A}$ is the input impedance of the antenna and $\rm \Gamma_{A}$ is the reflection coefficient of the antenna, again assuming a $\rm Z_{0}$ = 50~$\rm \Omega$  impedance for the measuring instrument.
  \item $\rm G$ is the power gain of the front-end amplifier in the radiometer.
  \item $\rm V_A$ is the voltage from the antenna terminals that is coupled into the transmission line, which is assumed to be of $\rm Z_{0}$ = 50~$\rm \Omega$ impedance.  $\rm V_A$ is a voltage waveform in the transmission line at the LNA input resulting from the coupling of antenna temperature to the line.
  \item $\rm V_N$ is the voltage of the noise wave generated in the LNA, referred to the amplifier input.
  \item $\rm f$ is the fraction of that noise wave voltage that gets coupled in the reverse direction into the $\rm Z_{0}$ = 50~$\rm \Omega$ transmission line connecting the antenna to the amplifier, which is of length $l$.
\end{itemize}
The amplifiers connected to the antenna in SARAS~3 are all in a compact module that is followed immediately by an optical modulator and hence is optically isolated from all electronics that follows.  Therefore, the total amplification---that of the first low-noise amplifier, a second amplification stage that follows, and the amplifier associated with the optical modulator---may be treated as lumped, referred to as the front-end amplifier, and represented by a single noise wave $\rm V_N$.  The analysis parameters are depicted in Fig.~\ref{fig:rad_mod}.

Taking into account the first order reflections of front-end amplifier noise from the antenna, the voltage $\rm V_S$ at the input of the amplifier can be written as
\begin{align}
\label{eq:V_S}
\rm V_S &= \rm  V_{A} (1+\Gamma_N) + V_{N}(1+\Gamma_N) + fV_N\Gamma_{A}\mathrm{e}^{i \rm \phi} (1+\Gamma_N)  \nonumber \\
&= \rm V_{A} (1+\Gamma_N) + V_{N} (1+\Gamma_N) [ 1+ f\Gamma_{A}\mathrm{e}^{i \rm \phi} ],
\end{align}
where  $\phi$ is the phase difference between the forward propagating wave and the reflected wave due to the finite  length $l$ of the transmission line connecting the amplifier and the antenna. $\phi$ and $ l$ are related as $\rm \phi = \rm (4\pi\nu {\it l})/(v_fc)$ where $c$ is the speed of light in vacuum and $v_f$ is the velocity factor of the transmission line. 

Taking into account the power gain of the amplifier, the time-averaged power flow out of the front-end amplifier is
%\begin{align}
\begin{eqnarray} \nonumber
\rm P_S & = & \rm \left\langle G~Re\Big(\frac{V_S V_S^*}{Z_N} \Big) \right\rangle \nonumber \\
\rm & = & \rm G~Re \{P_A [1-2iIm(\Gamma_N)-|\Gamma_N|^2]  \nonumber \\
\rm && \rm + P_N [1-2iIm(\Gamma_N)-|\Gamma_N|^2] [ 1+ f\Gamma_{A}\mathrm{e}^{i \rm \phi} ]  [ 1+ f^*\Gamma_{A}^*\mathrm{e}^{-i \rm \phi} ] \} \nonumber \\
\rm & = & \rm G~ (1-|\Gamma_N|^2)\{P_A + P_N  [1 + 2Re(f\Gamma_A\mathrm{e}^{i \rm \phi} ) + |f|^2|\Gamma_A|^2] \},
\label{eq:P_S_equ}
\end{eqnarray}
%\end{align}
where the following definitions are used
\begin{align}
\label{eq: P_A}
\rm P_A = \rm \left\langle \frac{V_A V_A^*}{Z_0}  \right\rangle
\end{align}
and
\begin{align}
\label{eq: P_N}
\rm P_N = \rm \left\langle \frac{V_N V_N^*}{Z_0} \right\rangle.
\end{align}
The relation $\rm Z_N (1-\Gamma_N) = Z_0 (1+\Gamma_N)$ is used in the above derivation to express $\rm Z_N$ in terms of $\rm Z_0$.  Additionally, it may be noted here that $\rm P_A$ represents the available power from the antenna: the power corresponding to the antenna temperature that couples into the transmission line, with characteristic impedance $\rm Z_0$,  connecting to the receiver.  $\rm P_N$ corresponds to the receiver noise, referred to the input of the LNA.

The correlation receiver response contains unwanted additives as a result of coupling of any common mode self-generated RFI or noise into the two arms of the correlation receiver.  For example, the samplers on the digital receiver board that digitise the analog signals of the two arms would inevitable have common mode noise of the digital board, which results in an unwanted additive component in the response.  This additive is expected to be constant in time and we denote the net unwanted common mode response as $\rm P_{cm}$.  With this power included, the measurements in each of two switch states, $\rm P_{OBS00}$ and $\rm P_{OBS11}$, are:

\label{eq:P1_OBS00}
\begin{eqnarray} \nonumber
\rm P_{\rm OBS00} & = & \rm -G~ (1-|\Gamma_N|^2)\{P_A + P_N  [1 + 2Re(f\Gamma_A\mathrm{e}^{i \rm \phi} ) + |f|^2|\Gamma_A|^2] \} + P_{\rm corr}
\end{eqnarray}
and
\begin{eqnarray} \nonumber
\label{eq:P1_OBS11}
\rm P_{\rm OBS11} & = & \rm G~ (1-|\Gamma_N|^2)\{P_A + P_N  [1 + 2Re(f\Gamma_A\mathrm{e}^{i \rm \phi} ) + |f|^2|\Gamma_A|^2] \} + P_{\rm corr}.
\end{eqnarray}
Their difference $\rm P_{OBS}$ is:
\begin{eqnarray} \nonumber
\label{eq:P1_OBS}
\rm P_{\rm OBS} & = & \rm P_{\rm OBS11} - P_{\rm OBS00} \nonumber \\
& = & \rm 2 G~ (1-|\Gamma_N|^2)\{P_A + P_N  [1 + 2Re(f\Gamma_A\mathrm{e}^{i \rm \phi} ) + |f|^2|\Gamma_A|^2] \}.
\end{eqnarray}

\begin{figure}[h!]
  \begin{center}
    \begin{circuitikz}
      \draw (0,0) to[short, f=$l$] (2,0) to[amp] (4,0);
      \draw (2,0) to[short, f=$\rm f$] (0,0);
      \draw (2,0) node[label={above:$\rm \Gamma_{N}$}] {};
      \draw (2.9,0.4) node[label={above:$\rm G, V_N$}] {};
      \draw [thick] (-1.5,-0.25) rectangle (0,0.25) node[pos=0.5]{Attenuator};
      \draw [thick] (-4,-0.5) rectangle (-2.0,0.5) node[pos=0.5]{Noise Source};
      \draw (-2,0) to [short] (-1.5,0);
      \draw (0,0.25) node[label={above:$\rm \Gamma_{REF}=0$}]{};
     \end{circuitikz}
    \caption{Simplified noise model for the SARAS radiometer, when connected to the reference termination and calibration source.}
    \label{fig:rad_mod2}
  \end{center}
\end{figure}
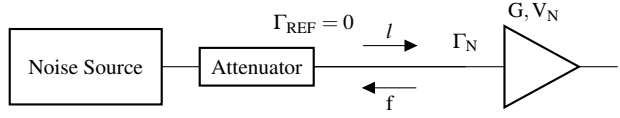

Consider the case in which, instead of an antenna, an impedance matched $\rm Z_0 = 50~\Omega$ calibration noise source is connected. This also serves as an ambient temperature reference termination when the noise source is off.  As there is no mismatch between the transmission line and noise source or reference termination, the noise wave from the amplifier that is coupled into the transmission line is absorbed at the calibration noise/reference termination. The analysis parameters in this case are depicted in Fig.~\ref{fig:rad_mod2}.
For this case, we may write the time-averaged power flowing out of the front-end amplifier as 
\begin{align}
\label{eq:P50_equ}
\rm P^{\prime}_{\rm S} = \rm G ~(P_{REF}+ P_N)(1 - |\Gamma_N|^2)
\end{align}
when the noise source is off and
\begin{align}
\label{eq:P50_equ}
\rm P^{\prime\prime}_{\rm S} = \rm G ~(P_{CAL}+ P_N)(1 - |\Gamma_N|^2)
\end{align}
when the noise source is on.  In deriving this, we have used Eq.~\ref{eq:P_S_equ} and set $\rm \Gamma_A = 0$ and replaced $\rm P_A$ with  $\rm P_{REF}$ or $\rm P_{CAL}$.  $\rm P_{REF}$ represents the noise power from the reference termination that couples into the transmission line, and $\rm P_{CAL}$ that from the termination when the calibration source is on.  We may write:
\begin{align}
\label{eq: P_REF}
\rm P_{REF} = \rm \left\langle \frac{V_{REF} V_{REF}^*}{Z_0}  \right\rangle,
\end{align}
where $\rm V_{REF} $ is the noise voltage from the reference port, and 
\begin{align}
\label{eq: P_CAL}
\rm P_{CAL} = \rm \left\langle \frac{V_{CAL} V_{CAL}^*}{Z_0}  \right\rangle,
\end{align}
where $\rm V_{CAL} $ is the noise voltage from the reference port when the calibration noise source is on.

Taking into account the unwanted common-mode noise from the digital boards, which are inevitably added, the measurement data provided by the correlation receiver in each of the states CAL00, CAL01, CAL10 and CAL11 may be written as:
\begin{align}
\label{eq:P1_CAL00}
\rm P_{\rm CAL00} = \rm -G (P_{REF}+ P_N)(1 - |\Gamma_N|^2) + P_{\rm corr},
\end{align}
\begin{align}
\label{eq:P1_CAL01}
\rm P_{\rm CAL01} = \rm  G (P_{REF}+ P_N)(1 - |\Gamma_N|^2) + P_{\rm corr},
\end{align}
\begin{align}
\label{eq:P1_CAL10}
\rm P_{\rm CAL10} = \rm -G (P_{CAL}+ P_N)(1 - |\Gamma_N|^2) + P_{\rm corr}
\end{align}
and
\begin{align}
\label{eq:P1_CAL11}
\rm P_{\rm CAL11} = \rm G (P_{CAL}+ P_N)(1 - |\Gamma_N|^2) + P_{\rm corr}.
\end{align}
Differencing the measurements recorded in the two switch positions gives
\begin{align}
\label{eq:P1_CAL0}
\rm P_{\rm CAL0} &= \rm P_{\rm CAL01} - P_{\rm CAL00} \nonumber \\
&=\rm 2G (P_{REF}+ P_N)(1 - |\Gamma_N|^2)
\end{align}
and
\begin{align}
\label{eq:P1_CAL1}
\rm P_{\rm CAL1} &= \rm P_{\rm CAL11} - P_{\rm CAL10} \nonumber \\
&=\rm 2G (P_{CAL}+ P_N)(1 - |\Gamma_N|^2).
\end{align}

The correlation spectrometer thus provides three differenced measurements: $\rm P_{\rm OBS}$ corresponding to when the antenna is connected to the receiver, $\rm P_{\rm CAL0}$ when the reference is connected, and $\rm P_{\rm CAL1}$ when the calibration noise is on.  Together with $\rm T_{STEP}$, these yield a calibrated measurement of the antenna temperature: 
\begin{eqnarray} \nonumber
\label{eq:T1_calib_appendix}
\rm T_{\rm meas} & = & \rm \frac{P_{\rm OBS}- P_{\rm CAL0}}{P_{\rm CAL1}- P_{\rm CAL0}} T_{\rm STEP} \nonumber \\
&=& \rm \frac{T_{\rm STEP}  
\Big[ P_A-P_{REF} + P_N [2Re(f\Gamma_A\mathrm{e}^{i \rm \phi} ) + |f|^2|\Gamma_A|^2]\Big]}{
\Big[ (P_{CAL}-P_{REF}) \Big]}.
\end{eqnarray}
This Eq.~\ref{eq:T1_calib_appendix} may be written in the form 
\begin{eqnarray} \nonumber
\label{eq:T1_calib_2}
\rm T_{\rm meas} & = & \rm T_{\rm STEP}\Big[\frac{P_A-P_{REF}} {P_{CAL}-P_{REF} }\Big] + \nonumber \\
&& \rm T_{\rm STEP}\Big[ \frac{P_N}{P_{CAL}-P_{REF}} \times  \Big\{ 2|f||\Gamma_A|cos(\phi_f+\phi_A+\phi) + |f|^2|\Gamma_A|^2 \Big\} \Big],
\end{eqnarray}
where $\phi_f$ is the phase associated with the complex $f$ and $\phi_A$ is the phase associated with the scattering parameter S11 of the antenna.  

So far, we have considered only first order reflection of the LNA noise from the antenna, which introduces sinusoidal standing waves with a single period within the transmission line and, consequently, sinusoidal modulation of the measured spectrum with a single period.  However, reflections of the LNA noise as well as the antenna signal that occurs at the input of the LNAs leads to higher order reflections and standing waves in the transmission line. We now proceed to quantify these reflections and associated spectral structure.

We begin with Eq.\ref{eq:V_S} and introduce higher order reflection terms. For clarity, we split the voltage at the input of the LNA into two parts, a part originating in the antenna and a second part corresponding to the LNA noise, and superpose the responses to get the resultant. Since the antenna signal and noise from the LNA are uncorrelated, this separation can be extended to the powers as well. 

The voltage due to the antenna, denoted as $\rm V_{SA}$, may be written as  
\begin{align}
\label{eq:V_S_A_2}
\rm V_{SA} &= \rm V_A (1+\Gamma_N) +V_A(\Gamma_N \Gamma_{A}\mathrm{e}^{i \rm \phi}) (1+\Gamma_N) +V_A(\Gamma_N^2 \Gamma_{A}^2\mathrm{e}^{i \rm 2\phi}) (1+\Gamma_N) .... \\
 &\rm ~~~~ + V_A(\Gamma_N^n \Gamma_{A}^n\mathrm{e}^{i \rm n\phi}) (1+\Gamma_N) + ....\nonumber \\
&=  \rm V_A \sum_{n=0}^{+\infty}(\Gamma_N\Gamma_{A}\mathrm{e}^{i \rm\phi})^n(1+\Gamma_N).
\end{align}
The time averaged power flow out of the system, due to the signal from the antenna, can be written as 
\begin{align} \nonumber
\label{eq:P_S_A_equ}
\rm P_{SA} & = \rm \left\langle G~Re\Big(\frac{V_{SA} V_{SA}^*}{Z_N} \Big) \right\rangle \nonumber \\
\rm & = \rm \left\langle G~Re\Big[ \frac{V_AV_A^*(1+\Gamma_N)(1+\Gamma_N^*) \{ \sum_{m=0}^{+\infty}(\Gamma_N\Gamma_{A}\mathrm{e}^{i \rm\phi})^m\} \{ \sum_{n=0}^{+\infty}(\Gamma_N^*\Gamma_{A}^*\mathrm{e}^{-i \rm\phi})^n\} } {Z_N} \Big] \right\rangle.
\end{align}
Using Cauchy product to evaluate the product of the two infinite series, the above expression may be simplified to 
 \begin{align} \nonumber
\rm P_{SA}  = & \rm G~Re\Big[ P_A \{1-2iIm(\Gamma_N)-|\Gamma_N|^2\} \{\sum_{k=0}^{+\infty}~|\Gamma_N|^k |\Gamma_A|^k\}\{ \sum_{l=0}^{k} e^{i(2l-k)(\phi_N+\phi_A+\phi)}\} \Big] \nonumber \\
= & \rm G~P_A \sum_{k=0}^{+\infty}~|\Gamma_N|^k |\Gamma_A|^k \sum_{l=0}^{k} \Big[cos\{(2l-k)(\phi_N+\phi_A+\phi)\}(1-|\Gamma_N|^2) \nonumber \\
& \rm ~~~~~~~~~~~~~~~~~~~~~~~~~~~~~~~~~~~~~+2Im(\Gamma_N)sin\{(2l-k)(\phi_N+\phi_A+\phi)\}\Big].
\label{eq:P_S_A_equ_2}
\end{align}
Since the last term containing the sine function is anti-symmetric, the summation of all of the sine terms is zero. Therefore, the expression for the power corresponding to the antenna becomes
\begin{align} 
\label{eq:P_SA_3}
\rm P_{SA} = & \rm G~P_A (1-|\Gamma_N|^2) \sum_{k=0}^{+\infty}~|\Gamma_N|^k |\Gamma_A|^k \sum_{l=0}^{k} cos\{(2l-k)(\phi_N+\phi_A+\phi)\}.
\end{align}

In a similar fashion, we may derive the voltage and power for the additive noise from the front-end amplifier. The voltage originating in the LNA is
\begin{align}
\label{eq:V_SL_1}
\rm V_{SN} &= \rm V_N(1+\Gamma_N) + fV_N\Gamma_A\mathrm{e}^{i \rm \phi}(1+\Gamma_N) +fV_N\Gamma_A^2 \Gamma_{N}\mathrm{e}^{i \rm 2\phi} (1+\Gamma_N) .... \\
 &\rm ~~~~ + fV_N\Gamma_N^{n-1} \Gamma_{A}^n\mathrm{e}^{i \rm n\phi} (1+\Gamma_N) + ....\nonumber \\
&=  \rm V_N(1+\Gamma_N) \big\{1 +f \sum_{n=0}^{+\infty}(\Gamma_A^{n+1}\Gamma_N^{n}e^{i(n+1)\phi})\big\}.
\end{align}
The power due to this voltage is given as 
\begin{align} \nonumber
\label{eq:P_S_N_equ}
\rm P_{SN} & = \rm \left\langle G~Re\Big(\frac{V_{SN} V_{SN}^*}{Z_N} \Big) \right\rangle \nonumber \\
\rm & = \rm G~Re\Big[ \frac{V_N V_N^*}{Z_N} (1+\Gamma_N)(1+\Gamma_N^*)\big\{1 + \sum_{m=0}^{+\infty}(f~\Gamma_A^{(m+1)}\Gamma_N^{m}e^{i(m+1)\phi})\big\} \\ \nonumber 
& \rm ~~~~~~~~~~~~~~~~~~~~~~~~~~ \times \big\{1+ \sum_{n=0}^{+\infty}(f^*~\Gamma_A^{*(n+1)}\Gamma_N^{*n}e^{-i(n+1)\phi})\big\} \Big] \nonumber \\
\rm & = \rm G~Re\Big[P_N \big\{1-2iIm(\Gamma_N)-|\Gamma_N|^2\big\} \big\{1 + \sum_{m=0}^{+\infty}(f~\Gamma_A^{(m+1)}\Gamma_N^{m}e^{i(m+1)\phi})\big\} \\ \nonumber
& \rm  ~~~~~~~~~~~~~~~~~~~~~~~~~~ \times  \big\{1+ \sum_{n=0}^{+\infty}(f^*~\Gamma_A^{*(n+1)}\Gamma_N^{*n}e^{-i(n+1)\phi})\big\} \Big].
\end{align}
The above equation simplifies to
\begin{align}
\label{eq:P_SN_equ2}
\rm P_{SN}  &= \rm G~P_N ~Re\Big[\big\{1-2iIm(\Gamma_N)-|\Gamma_N|^2\big\}  \big\{1 + \rm \sum_{m=0}^{+\infty}2Re(f~\Gamma_A^{m+1}\Gamma_N^{m}e^{i(m+1)\phi}) + \\ \nonumber 
&~~~~~~~~~~~~~~~~~~~~~~~~~~  \rm  |f|^2|\Gamma_A|^2\sum_{k=0}^{+\infty}|\Gamma_N|^k|\Gamma_A|^k\sum_{l=0}^{k} e^{i(2l-k)(\phi_N+\phi_A+\phi) } \big\} \Big].
\end{align}

Expanding terms, identifying $\rm m=k$, and using arguments similar to that used in derivations above for the case of a single reflection at the antenna, we obtain
\begin{align} 
\label{eq:P_SN_3}
\rm P_{SN} = \rm G~P_N (1-|\Gamma_N|^2)\Big[1 + & \rm \sum_{k=0}^{+\infty} (2|f||\Gamma_A|^{(k+1)}|\Gamma_N|^k cos\{\phi_f +(k+1)(\phi_A+\phi) +k\phi_N\}) \\ \nonumber
+ &\rm |f|^2|\Gamma_A|^2\sum_{k=0}^{+\infty}~|\Gamma_N|^k |\Gamma_A|^k \sum_{l=0}^{k} cos\{(2l-k)(\phi_N+\phi_A+\phi)\} \Big].
\end{align}

The total power flow out of the system can then be expressed as 

\begin{align} 
\label{eq:P_S_final} 
\rm P_{S} & = \rm P_{SA} + P_{SN} \\ \nonumber
 & \rm =  G(1-|\Gamma_N|^2)  \Big[ P_N\big[1 +  \rm \sum_{k=0}^{+\infty} (2|f||\Gamma_A|^{(k+1)}|\Gamma_N|^k cos\{\phi_f +(k+1)(\phi_A+\phi) +k\phi_N\}) \\ \nonumber
& ~~~~~~~~~~~~~~~~~~~~~~~~~~ +  \rm |f|^2|\Gamma_A|^2\sum_{k=0}^{+\infty}~|\Gamma_N|^k |\Gamma_A|^k \sum_{l=0}^{k} cos\{(2l-k)(\phi_N+\phi_A+\phi)\} \big] \\ \nonumber
& ~~~~~~~~~~~~~~~~~~~~~~~~~~ +  \rm P_A \sum_{k=0}^{+\infty}~|\Gamma_N|^k |\Gamma_A|^k \sum_{l=0}^{k} cos\{(2l-k)(\phi_N+\phi_A+\phi)\} \Big]
\end{align}

It may be noted here that equations~\ref{eq:P1_CAL0} and \ref{eq:P1_CAL1} for the calibration states remain unchanged since it is assumed here that the reference port is impedance matched to the transmission line and both have impedances $\rm Z_0$; there are no reflections of voltage waveforms at the reference port. 

Omitting the pedagogical steps, the calibrated spectrum may thus be written as

\begin{align}
\label{eq:Tinf_calib}
\rm T_{\rm meas} & = \rm T_{\rm STEP}\Big\{ \rm  \frac{P_A [ \sum_{k=0}^{+\infty}~|\Gamma_N|^k |\Gamma_A|^k \sum_{l=0}^{k} cos\{(2l-k)(\phi_N+\phi_A+\phi)\}]-P_{REF}} {P_{CAL}-P_{REF} } \\ \nonumber
& + \rm \frac{P_N}{P_{CAL}-P_{REF}} \times \Big[\sum_{k=0}^{+\infty} (2|f||\Gamma_A|^{(k+1)}|\Gamma_N|^k cos\{\phi_f +(k+1)(\phi_A+\phi) +k\phi_N\}) \\ \nonumber
&~~~~~~~~~~~~~~~~~~~~~~~~~~~ +  \rm |f|^2|\Gamma_A|^2\sum_{k=0}^{+\infty}~|\Gamma_N|^k |\Gamma_A|^k \sum_{l=0}^{k} cos\{(2l-k)(\phi_N+\phi_A+\phi)\} \Big]\Big\}.
\end{align}

If we set $\rm k=0$ in the above equation, we recover Equation~\ref{eq:T1_calib_2} that represents the measured temperature assuming single reflection at the antenna and neglecting higher order terms.

\end{appendices}
\newpage
\bibliographystyle{spmpsci}       
\bibliography{references}  
\end{document}